\documentclass[fleqn,usenatbib]{mnras}

\usepackage{newtxtext,newtxmath}
\usepackage{graphicx}
\usepackage{amsmath}
\usepackage{xfrac}
\usepackage{subfig}

\usepackage[T1]{fontenc}
\usepackage{ae,aecompl}

\usepackage{graphicx}	
\usepackage{amsmath}	
\usepackage{amssymb}	
\usepackage{threeparttable}
\usepackage{float}
\usepackage{hhline}
\usepackage{psfrag}

\newcommand{\msun}{M_{\odot}}

\newcommand{\erf}{\mathrm{erf}\,}
\newcommand{\thetahat}{\hat{\theta}}

\title[Characterizing Galaxy Velocity Fields using the Radon Transform]{SDSS IV MaNGA: Characterizing Non-Axisymmetric Motions in Galaxy Velocity Fields Using the Radon Transform}

\author[D.~V. Stark et al.]{David V. Stark,$^{1}$\thanks{E-mail:david.stark@ipmu.jp}
Kevin A. Bundy,$^{2}$ Kyle Westfall,$^{2}$ Matt Bershady,$^{3}$ 
\newauthor Anne-Marie Weijmans,$^{4}$ Karen L. Masters,$^{5,6}$ Sandor Kruk,$^{7}$ Jarle Brinchmann,$^{8}$ \newauthor Juan Soler,$^{9}$  Roberto Abraham,$^{10}$ Edmond Cheung,$^{1}$ Dmitry Bizyaev,$^{11,12}$ Niv Drory,$^{13}$ \newauthor Alexandre Roman Lopes$^{14},$ David R. Law$^{15}$
\\
$^{1}$Kavli IPMU (WPI),The University of Tokyo Institutes for Advanced Study, The University of Tokyo, Kashiwa, Chiba 277-8583, Japan\\
$^{2}$UCO/Lick Observatory, University of California, Santa Cruz, 1156 High St. Santa Cruz, CA 95064, USA\\
$^{3}$Department of Astronomy, University of Wisconsin-Madison, 475N. Charter St., Madison WI 53703, USA	\\
$^{4}$School of Physics and Astronomy, University of St Andrews, North Haugh, St. Andrews KY16 9SS, UK	\\
$^{5}$ Haverford College, Department of Physics and Astronomy, 370 Lancaster Avenue, Haverford, Pennsylvania 19041, US \\
$^{6}$University of Portsmouth, Institute of Cosmology \& Gravitation, Dennis Sciama Building, Portsmouth, PO1 3FX, UK \\
$^{7}$ Sub-department of Astrophysics, Department of Physics, University of Oxford, Denys Wilkinson Building, Keble Road, Oxford OX1 3RH\\
$^{8}$Leiden Observatory, Leiden University, PO Box 9513, 2300 RA Leiden, the Netherlands \\
$^{9}$Institut d'Astrophysique Spatiale, Universit{\'e} Paris-XI, Orsay, France  \\
$^{10}$Department of Astronomy and Astrophysics, University of Toronto, 50 St. George Street, Toronto, ON M5S 3H4, Canada\\
$^{11}$Apache Point Observatory, P.O. Box 59, Sunspot, NM 88349 \\
$^{12}$Special Astrophysical Observatory of the RAS, 369167, Nizhnij Arkhyz, Russia\\
$^{13}$McDonald Observatory, The University of Texas at Austin, 1 University Station, Austin, TX 78712, USA	\\
$^{14}$Departamento de Fisica, Facultad de Ciencias, Universidad de La Serena, Cisternas 1200, La Serena, Chile \\
$^{15}$Space Telescope Science Institute, 3700 San Martin Drive, Baltimore, MD 21218, USA
}
\date{Accepted 2018 July 24. Received 2018 July 21; in original form 2018 May 25}

\pubyear{2018}

\begin{document}
\label{firstpage}
\pagerange{\pageref{firstpage}--\pageref{lastpage}}
\maketitle

\begin{abstract}
We show how the Radon transform {(defined as a series of line integrals through an image at different orientations and offsets from the origin)} can be used as a simple, non-parametric tool to characterize galaxy velocity fields, specifically their global kinematic position angles (PA$_k$) and any radial variation or asymmetry in PA$_k$.  This method is fast and easily automated, making it particularly beneficial in an era where IFU and interferometric surveys are yielding samples of thousands of galaxies. We demonstrate the Radon transform by applying it to gas and stellar velocity fields from the first $\sim$2800 galaxies of the SDSS-IV MaNGA IFU survey. We separately classify gas and stellar velocity fields into five categories based on the shape of their radial  PA$_k$ profiles.  At least half of stellar velocity fields and two-thirds of gas velocity fields are found to show detectable deviations from uniform coplanar circular motion, although most of these variations are symmetric about the center of the galaxy. The behavior of gas and stellar velocity fields is largely independent, even when PA$_k$ profiles for both components are measured over the same radii. We present evidence that one class of symmetric PA$_k$ variations is likely associated with bars and/or oval distortions, while another class is more consistent with warped disks.  This analysis sets the stage for more in-depth future studies which explore the origin of diverse kinematic behavior in the galaxy population.
\\
\\
\end{abstract}


\pagebreak



\section{Introduction}

Galaxy kinematics provide a powerful means of understanding the physical processes that govern galaxy evolution.  In particular, deviations from coplanar circular motion potentially reveal the existence of non-uniformities in galaxies' matter distributions, inflows or outflows, and/or tidal interactions.  In many cases, the origin of certain kinematic irregularities is still debated.

Deviations from simple rotation are commonly observed throughout the galaxy population, and come in a variety of different forms.  For instance, anywhere from 20--50\% of disk galaxies have detectable asymmetries in their rotation curves or projected 2D velocity fields \citep{Haynes98,Swaters99,Kornreich00,Kannappan02,Andersen13,Bloom17}.  Another frequent phenomenon seen in both gas and stellar disks are ``warps", where the kinematic position angle in the outer disk is different from that of the inner disk, implying the presence of a disk with a radially varying inclination \citep{Sancisi76,Bosma81a,Bosma81b,Garcia-Ruiz02,Reshetnikov02,Schwarzkopf01,Ann06}. Warps frequently begin at or just beyond the optical radius $R_{25}$, the radius where the $B$ band surface brightness reaches $25\,{\rm mag\,arcsec^{-2}}$ \citep{Briggs90, Ann06,VanDerKruit07}, although they can begin at smaller radii \citep{VanDeVoort15,Reshetnikov16}.  The high frequency of asymmetric and warped disks suggests that these features are either frequently generated or long lived. Although at least some fraction of them are likely due to tidal interactions \citep{Ann06}, they are also commonly seen in in low density environments, suggesting additional physical drivers.  Gas and/or dark matter infall may provide another explanation \citep{Ostriker89,Jiang99,Bournaud05,Shen06,VanDeVoort15}, particularly in low density environments which are more likely to host gas-rich mergers and ``cold-mode" cosmological accretion \citep[e.g.][]{Keres09}.  

Inner regions of galaxies (well within $R_{25}$) can also show kinematics that are distinct from the rest of the galaxy. Bar instabilities, which follow solid body rotation not necessarily aligned with the major axis of the galaxy disk, are thought to occur in at least $\sim$30\% of galaxies \citep{Masters11}. The individual stars within bars follow highly elliptical orbits, but bars can also drive radial motions within the gas.  Similar behavior is seen in ``oval distortions" (the distinction between bars and oval distortions in the literature appears somewhat subjective, but we consider them essentially less extreme bars). A commonly observed phenomenon, particularly in early type galaxies, are ``kinematically-decoupled cores'' (KDCs), where the motions within the inner few kpc are misaligned with the rest of the galaxy  \citep{Bender88,Franx88,Davies01,McDermid06,Emsellem07,Krajnovic11}. KDCs may be formed by major mergers, or by star formation in recently acquired misaligned gas  \citep{Kormendy84,HolleyBockelmann00,Balcells90, Hernquist91,Bois10, Bois11, Tsatsi15}, but may in some cases be a projection effect of different orbit families \citep{VandenBosch08}.

Observationally constraining the primary drivers of various types of {non-axisymmetric motions} would benefit from large samples of galaxies with measured velocity fields that also span a wide range of properties, allowing a reliable estimate of the frequency of different kinematic features and how they relate to other galaxy characteristics.  The MaNGA (Mapping Nearby Galaxies at Apache Point Observatory; \citealt{Bundy15}) survey is currently the largest integral field unit (IFU) survey in existence, making it an ideal data set to conduct a statistical study of galaxy kinematics in the $z=0$ universe. Additionally, the rapid growth of IFU survey sample sizes has created new demand for analysis techniques which can reliably characterize velocity fields with minimal human supervision.

With these goals in mind, we have developed a method to quantify the radial variation in the kinematic position angles (PA$_k$) of galaxies based on the Radon transform, which can then be used to identify deviations from simple co-planar rotation in velocity fields. We then demonstrate this method on data from the SDSS IV MaNGA survey.  We identify several different characteristic patterns in the way kinematic position angles vary within galaxies, and use these patterns to classify galaxies into five distinct categories.  We then examine several basic properties of these different types, including their frequency, structural properties, agreement between gas and stellar velocity fields, color-mass distribution, and whether they host bars. Our demonstration of this new method of characterizing galaxy position angles and the subsequent analysis sets the stage for more detailed studies of disks with irregular kinematics to be carried out in the future.  

\section{The Radon Transform}
\label{sec:Radon}

\begin{figure}
	\includegraphics[width=\columnwidth]{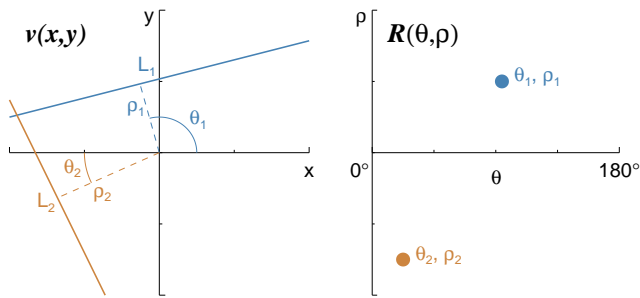}
    \caption{Illustration of the Radon transform and its coordinate system. Integrals are calculated along all possible lines, parameterized by the coordinates [$\theta$,$\rho$], that cross the 2D function $v(x,y)$.  Two examples are shown in the left panel, where the integrals are calculated over the solid lines, $L_1$ and $L_2$, which are {\it perpendicular} to the [$\theta$,$\rho$] vectors.  The values of these integrals are then plotted in [$\theta$,$\rho$] parameter space (right panel).  Under this coordinate system, $\theta$ ranges from 0--180$^{\circ}$ while $\rho$ ranges from $-\infty$ to $\infty$ such that a position below the $x$-axis corresponds to $\rho<0$.}
    \label{fig:Radon_diagram}
\end{figure}

Our method of analyzing velocity fields is based on the Radon transform, $R$ \citep{Radon17}:
\begin{equation}
\label{eq:Radon}
R(\rho,\theta)=\int_L{v(x,y) dl}
\end{equation}
where $v(x,y)$ is a 2D function (in our case, a velocity field\footnote{See also \citet{Starck03}, \citet{Case09}, and \citet{Krone-Martins13} for other recent applications of the Radon transform to astronomical data.}) and the subscript $L$ denotes a line integral. $R$ is a transform whereby integrals are calculated along lines that cross $v(x,y)$ at different orientations and distances from the origin.  These lines are parameterized by the polar coordinates in the plane of the sky $\left[\theta,\rho\right]$ where $\theta$ is the angle with respect to the $x$-axis and $\rho$ is the distance from the origin. Each integral is calculated along the line {\it perpendicular} to the [$\theta,\rho$] vector (see Fig.~\ref{fig:Radon_diagram}).  In the output coordinate system, $\theta$ spans from $0^{\circ}$ to $180^{\circ}$ while $\rho$ spans from $-\infty$ to $+\infty$, so any regions below the $x$-axis are considered to have $\rho<0$. The sign on the $\rho$ vector allows asymmetries in $R$ to be easily identified (Section~\ref{sec:characteristic_output}.

\begin{figure*}
	\includegraphics[width=2\columnwidth]{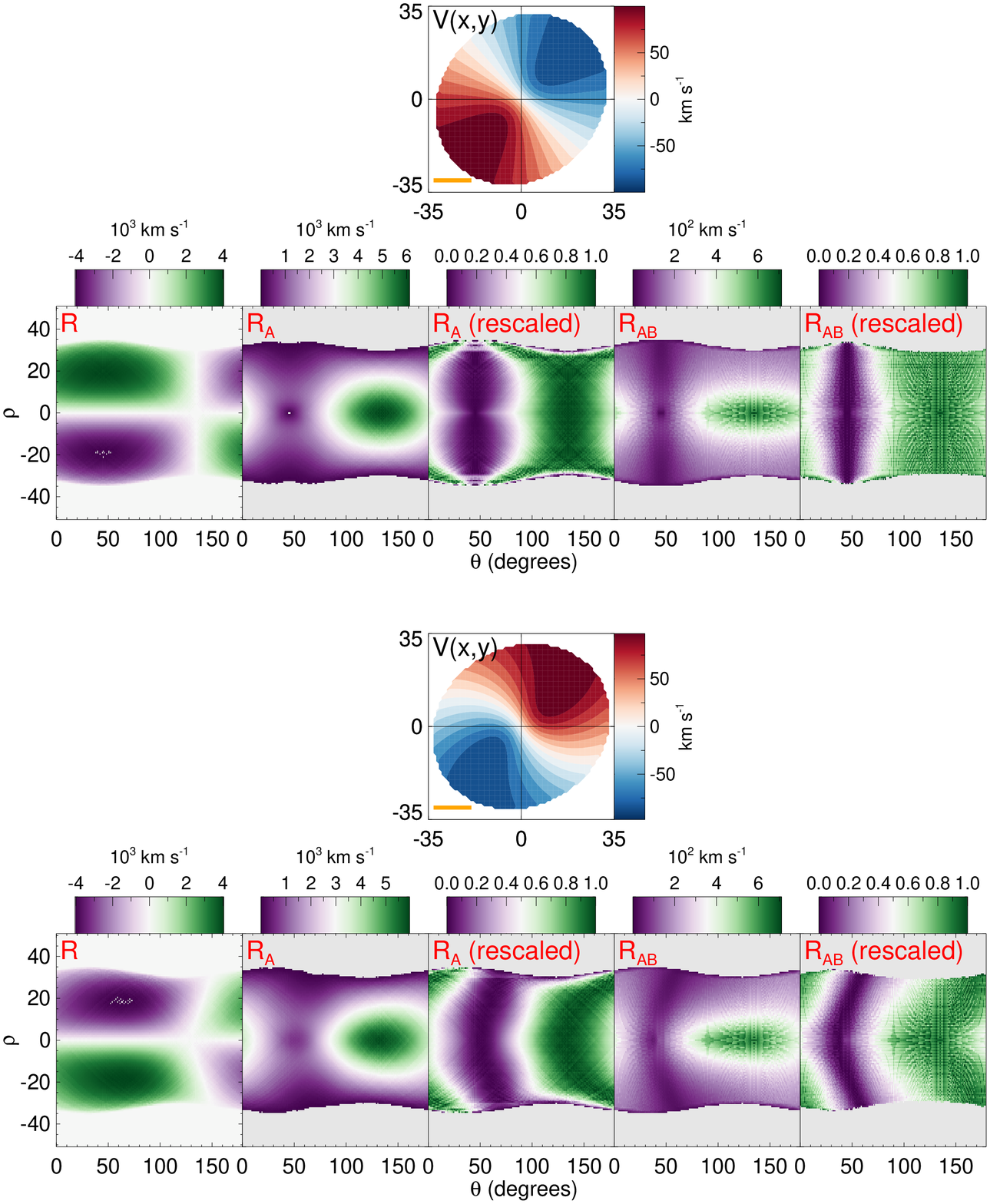}
    \caption{Demonstration of the Radon transform applied to two simple model velocity fields, one a normal rotating disk (top) and one a warped rotating disk (bottom). The model velocity field is of the form $v=v_0\tanh(r/h)\sin(i)\cos(\phi-PA_k)$, where $r$ is the radius, $i$ is the inclination, $\phi$ is the angle with respect to the positive $y$ axis, and $v_0$ and $h$ are constants that define the true rotation velocity.  We set $v_0=200\,{\rm km\,s^{-1}}$, $h=7$ pixels, $i=20^{\circ}$. The top velocity field has PA$_k$=135$^\circ$ (defined as the angle north through east of the receding major axis) while the bottom velocity field has a PA$_k$ approximately $180^{\circ}$ larger. The maximum radius is set to 2.5$R_e$ (where $R_e$ is the effective radius), and we set $R_e=2h$. For the warped velocity field, PA$_k$ changes with radius at a rate of $15^{\circ}\,R_e^{-1}$.  Below each velocity field, the panels show (from left to right): the Radon transform ($R$), the Absolute Radon transform ($R_A$), a rescaled version of $R_A$ where values at fixed $\rho$ span from 0--1 (see text), the Absolute Bounded Radon transform ($R_{\rm AB}$), and a rescaled version of $R_{\rm AB}$.  The thick orange lines to the bottom-left of each velocity field shows the size of the ``radon aperture" ($2\times r_{\rm ap}$) used to calculate $R_{\rm AB}$. }
    \label{fig:Radon_transform_example}
\end{figure*}

We apply a simple modification to the Radon transform that provides more useful information about the orientation of galaxy velocity fields. Instead of integrating over the raw velocity measurements, we instead integrate the absolute value of the difference between each point $v(x_i,y_i)$ and the mean of all values along each line segment:
\begin{equation}
\label{eq:absolute_Radon}
R_A=\int{|v(x,y)-\left\langle v(x,y)\right\rangle| dl}
\end{equation}
This modified transform is referred to as the {\it Absolute} Radon Transform, $R_A$, which reflects the amount of change in velocity along each line segment without having to directly calculate derivatives.  Examples of $R$ and $R_A$ calculated for different model velocity fields are shown in Fig.~\ref{fig:Radon_transform_example}.  

\begin{figure}
\includegraphics[width=\columnwidth]{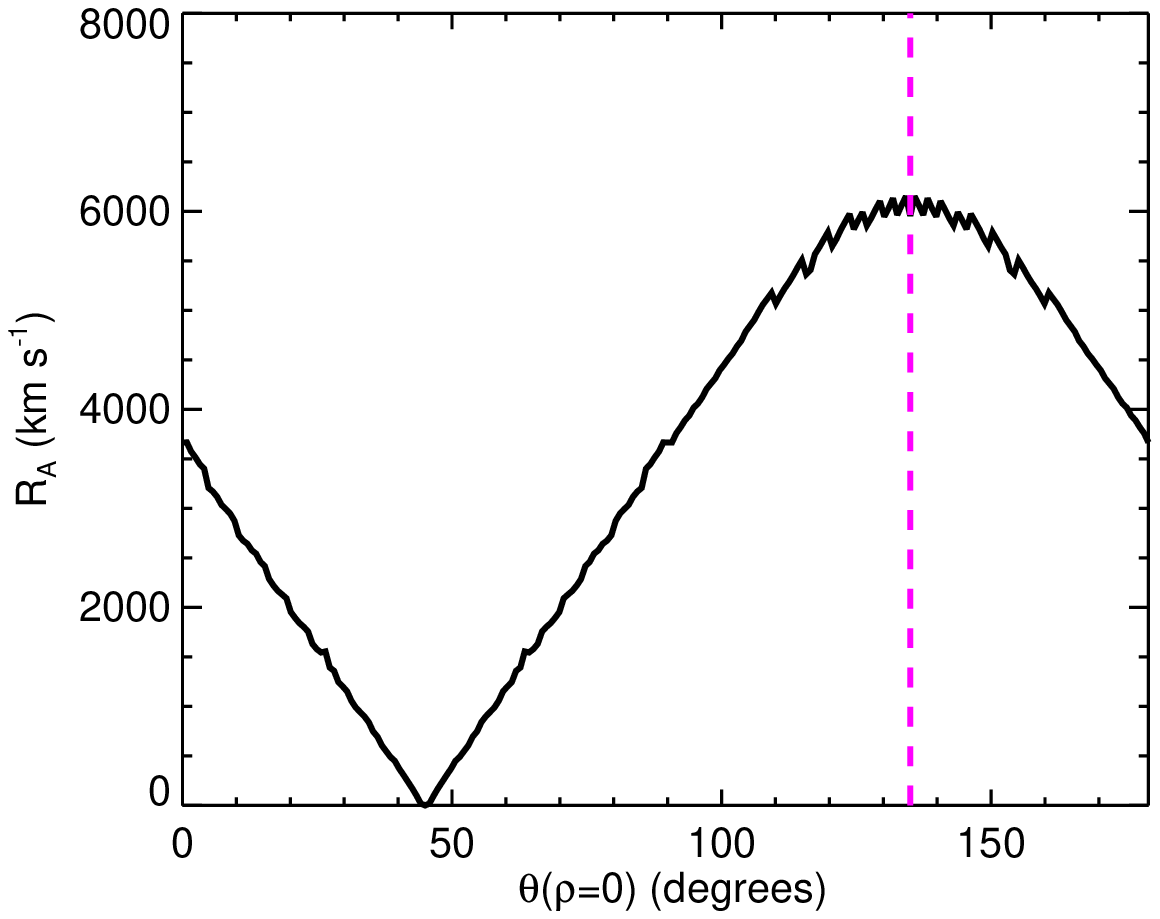}
\caption{Slice through $R_A$ from the top panel of Fig.~\ref{fig:Radon_transform_example} at $\rho=0$. The vertical dashed line indicates the true PA$_k$ of the model, which corresponds to where $R_A$ is maximized.}
\label{fig:Radon_pa_calc}
\end{figure}

A simple application of $R_A$ is to estimate the mean PA$_k$ of a velocity field, which should correspond to the line segment crossing through the center of the galaxy ($\rho=0$) at the angle $\theta$ where $R_A$ is maximized. This application is illustrated in Fig.~\ref{fig:Radon_pa_calc} which plots a slice of $R_A$ taken from Fig.~\ref{fig:Radon_transform_example} at $\rho=0$. The location where $R_A$ is maximized is in good agreement with the expected value (vertical dashed line) given the true PA$_k$. It is important to note that $R_A$ does not distinguish between the approaching/receding sides of the velocity field, but this distinction may be useful in some situations, such as when one wants to identify counter-rotating gas and stellar disks. However, the standard Radon transform $R$ is sensitive to whether velocities are positive or negative (as seen in the maps of $R$ in Fig.~\ref{fig:Radon_transform_example} where the two galaxies rotate in opposite directions), and can be easily used to infer the direction of the approaching and receding sides of a velocity field. 

We remind the reader that the definition of angles in the Radon transform is different from how angles are typically defined in astronomical data. For instance the galaxy PA$_k$ is traditionally defined with respect to the $y$-axis, while $\theta$ is defined with respect to the $x$-axis, and yet the value of $\theta$ where $R_A$ is maximized is equivalent to the PA$_k$.  To clarify why this is the case, we reiterate that $R_A$ is actually calculated along a line {\it perpendicular} to the [$\theta$,$\rho$] vector (Fig.~\ref{fig:Radon_diagram}). This $90^{\circ}$ difference makes it so the value of $\theta$ (defined relative to the $x$-axis) is equivalent to the PA$_k$ (defined relative to the $y$-axis).  

Although measuring the global PA$_k$ is a useful application of the Absolute Radon transform, our primary goal is to track radial variations in the PA$_k$ which can indicate deviations from {uniform co-planar} circular motion.  If we focus on where $R_{A}$ is {\it minimized} rather than where it is maximized, we find that $R_A$ shows an easily identifiable response to radial variations in PA$_k$.  We illustrate this behavior in Fig.~\ref{fig:Radon_transform_example} (3rd column) after offsetting and renormalizing $R_A$ so that each row ranges from 0--1, making the behavior at large $\rho$ more apparent.  In the top example of Fig.~\ref{fig:Radon_transform_example} where the galaxy has constant PA$_k$, the value of $\theta$ where $R_A(\rho)$ is minimized remains constant, while in the second example where we introduce a radial variation in PA, the value of $\theta$ where $R_A$ is minimized is clearly varying with $\rho$. Although we have highlighted the behavior where $R_A$ is minimized, the region where $R_A$ is maximized actually shows similar behavior in the presence of variations in PA$_k$.  However, after additional modifications to the Radon Transform (discussed below), it will become more clear why focusing on where $R_A$ is minimized is ideal.

One major issue with $R_A$ in its current form is that it will depend not only on the values of velocity measurements along each line segment, but the number of spaxels being integrated along a given line segment.  Thus $R_A$ runs the risk of reflecting the directions along which there is simply more or less data (more akin to a photometric position angle if the amount of available data is dependent on surface brightness), rather than directions where the velocity is truly changing by large or small amounts.  To solve this issue, we introduce integration bounds ($\pm r_{ap}$) on Eq.~\ref{eq:absolute_Radon}, where $r_{ap}$ (which we refer to as the {\it Radon aperture}) can be set to any value, but ideally one small enough such that the amount of data included in each line integral is independent of $\rho$ and $\theta$ (at least away from the edges of the velocity field). We refer to this {\it bounded} Absolute Radon transform as $R_{AB}$, but aside from the integration limits, the functional form is identical to $R_A$.  The final two columns in Fig.~\ref{fig:Radon_transform_example} illustrate $R_{AB}$ applied to model velocity fields.  In addition to being less impacted by varying numbers of spaxels in each line integral, the region where $R_{AB}$ is minimized is more sharply defined compared to where $R_A$ is minimized. As mentioned above, the region where $R_{\rm AB}$ is {\it maximized} also shows a clear response to radial variations in PA$_k$. However, the region where $R_{\rm AB}$ is minimized is sharper and thus a better indicator of PA$_k$. Furthermore, the band along which $R_{\rm AB}$ is minimized as a function of $\rho$ is tracing the kinematic major axis, whereas the band where $R_{\rm AB}$ is maximized is tracing the kinematic minor axis (see Fig.~\ref{fig:Radon_model_vary_aperture}). A more detailed discussion of how $R_{AB}$ depends on the choice of $r_{\rm ap}$ and our final choice of $r_{\rm ap}$ for our analysis of MaNGA data will be discussed in more detail in Section~\ref{sec:sys_apsize} and Section~\ref{sec:rap_choice}

It is important to remember that the relationship between $\theta$ and PA$_k$ depends on whether we are focusing on the value of $\theta$ where $R_{\rm AB}$ is minimized or maximized.  When we consider the value of $\theta$ where $R_{\rm AB}$ is {\it maximized}, $\theta$ matches the PA$_k$ defined in the astronomical convention.  However, when we consider the value of $\theta$ where $R_{\rm AB}$ is {\it minimized}, $\theta$ no longer matches the PA$_k$ but is offset by $90^{\circ}$.

\subsection{Radon Profile Measurement}
\label{sec:tracing_algorithm}

The key feature of $R_{\rm AB}$ is the ``ridge'' along which $R_{\rm AB}(\rho)$ is minimized (the dark purple region in the right panels of Fig.~\ref{fig:Radon_transform_example}). We refer to the values of $\theta$ where $R_{\rm AB}$ is minimized as $\thetahat(\rho)$, and this is one of the important measurements we make throughout this work as it is a direct indicator of PA$_k$ (with a $90^{\circ}$ offset). Throughout our analysis we create 1D profiles of $\thetahat(\rho)$ which we will hereafter refer to as {\it Radon profiles}. The following algorithm is used to extract Radon profiles from 2D maps of $R_{\rm AB}$:
\begin{enumerate}
\item We first flag any regions of $R_{\rm AB}$ where the estimated value may be biased because the line integral at that [$\theta$,$\rho$] overlaps missing/flagged spaxels or the edge of the velocity field.
\item Starting from $\rho=0$, we smooth $R_{\rm AB}(\theta,\rho=0)$ using a kernel with width equal to 15\% of the full range of $\theta$. We then identify all local minima and maxima, and take the strongest local minimum as our initial guess of $\thetahat$.  
\item We fit the {\it unsmoothed} $R_{\rm AB}$ vs. $\theta$ with a von Mises function (i.e., a Gaussian distribution for polar coordinate, but with negative amplitude since we are fitting where $R_{\rm AB}$ is minimized), and the centroid of this function is our estimate of $\thetahat$.  Because there can be secondary minima in $R_{\rm AB}$, we restrict our fit to only use data within the two local maxima nearest the initial guess of $\thetahat$, or $\pm 45^\circ$, whichever is closer.
\item We then iterate over $\rho>0$ and $\rho<0$ (from small to large $|\rho|$), repeating the steps above with a few exceptions: (a) Instead of using the location of the strongest minima in $R_{\rm AB}$ as our initial guess for $\thetahat$, we use the estimate of  $\thetahat$ from the previous value of $\rho$. (b) We restrict each $\thetahat$ to be within $\pm 30^{\circ}$ of the previous estimate.  
\item Any $\thetahat$ measurement whose 95\% confidence bounds overlap regions in $R_{\rm AB}$ with missing data are flagged and removed from any subsequent analysis. Our calculation of uncertainty when using real data is discussed in Section~\ref{sec:uncertainty}.
\end{enumerate}

\subsection{Systematic Errors}

We have illustrated how $R_{\rm AB}$ can be used to identify radial variations in PA$_k$.  In this section we highlight systematic errors which must be considered when applying our algorithm to velocity fields.

\subsubsection{Size of Radon Aperture ($r_{\rm ap}$)}
\label{sec:sys_apsize}

\begin{figure*}
	\includegraphics[width=2\columnwidth]{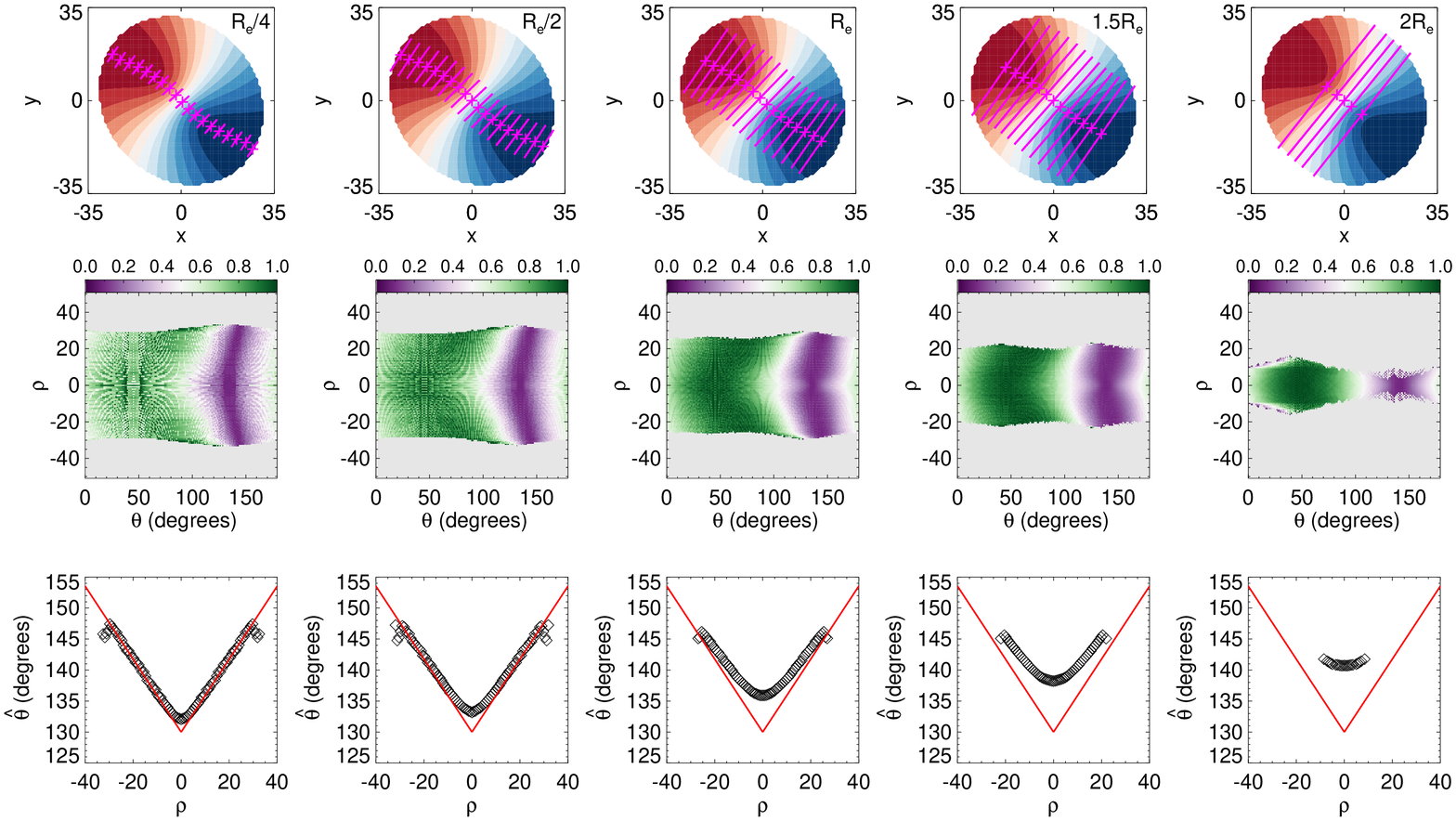}
    \caption{Model velocity fields (top), rescaled $R_{\rm AB}$ (middle), and Radon profiles $\thetahat(\rho)$ (bottom) for different choices of $r_{\rm ap}$ (indicated in the upper right corner of each panel in the top row), ranging from $R_e/4$ to $2R_e$. In this and all subsequent figures, we scale $r_{\rm ap}$ by the minor-to-major axis ratio (in this case $b/a=\cos{i}$) so that the line segment being integrated over is approximately equal to $r_{\rm ap}$ projected onto the face of the galaxy disk along but perpendicular to the expected major axis. The model velocity field is the same as in Fig.~\ref{fig:Radon_transform_example} except we use a different central PA$_k$ and also introduce a radial variation in the PA$_k$ of $10^{\circ}\,R_e^{-1}$. Magenta points in the top row trace the kinematic major axis based on $\thetahat$ and lines extending out from each of these points illustrate the size of $r_{\rm ap}$ used to calculate $R_{\rm AB}$ at that position. The red line in the bottom row indicates the expected value of $\thetahat$ based on true PA$_k$ as a function of radius.}
    \label{fig:Radon_model_vary_aperture}
	\includegraphics[width=2\columnwidth]{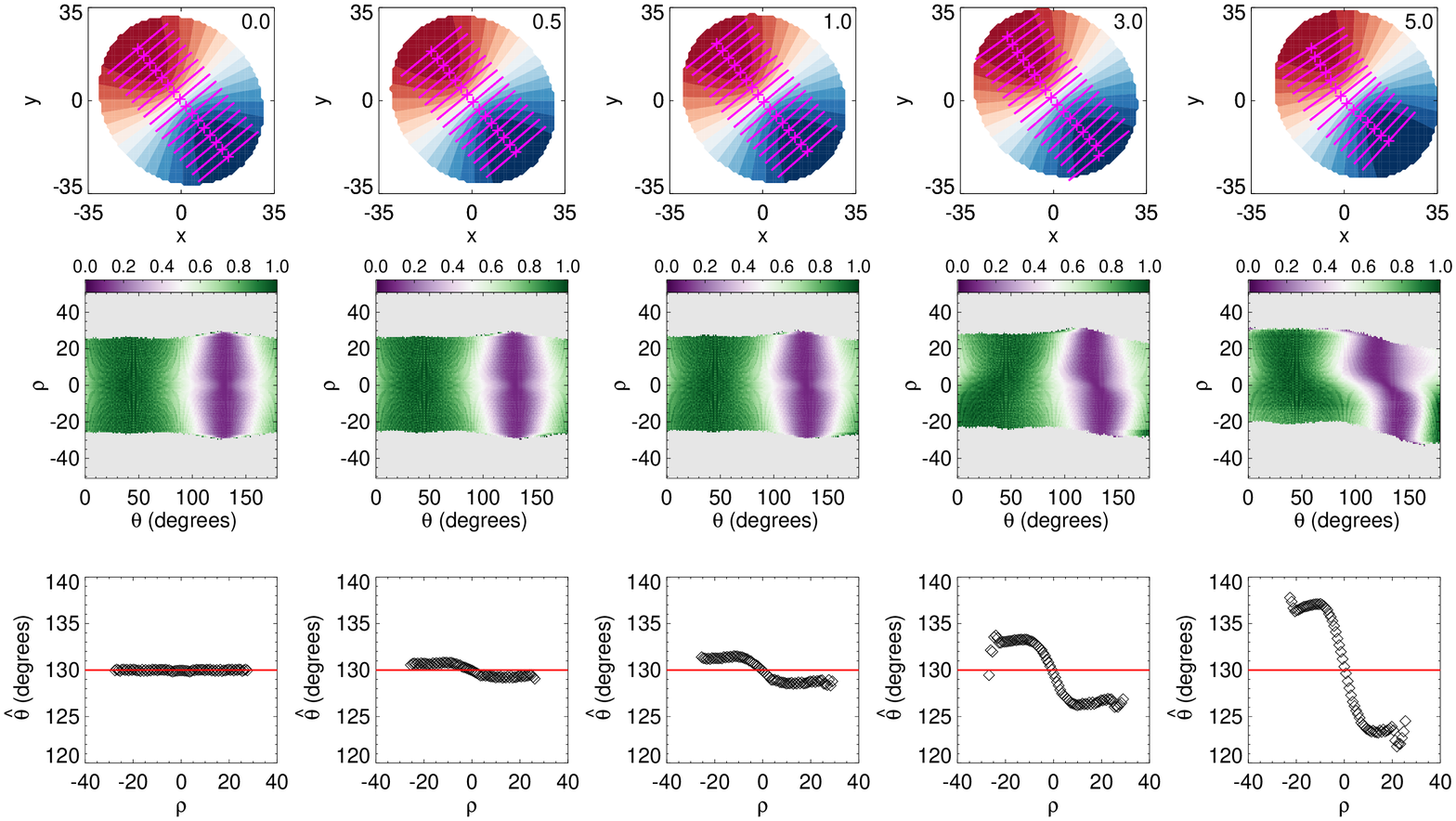}
    \caption{Same as Fig.~\ref{fig:Radon_model_vary_aperture} but showing the impact of shifting the velocity field off center by different amounts up to 5 spaxels (indicated to the upper-right of each velocity field).  The velocity field model is also the same as Fig.~\ref{fig:Radon_model_vary_aperture} but without any intrinsic variation in PA$_k$.}
    \label{fig:Radon_model_vary_offcenter}
\end{figure*}

In Section~\ref{sec:Radon} we introduced $r_{\rm ap}$ which defines the bounds of all the line integrals when calculating $R_{\rm AB}$.  One has the freedom to set any value of $r_{\rm ap}$, but there are pros and cons associated with different choices.  In Fig.~\ref{fig:Radon_model_vary_aperture}, we demonstrate how different choices for $r_{\rm ap}$ (ranging from $R_e/4$ to $2R_e$, where $R_e$ is the effective radius) can affect $R_{\rm AB}$ and the extracted Radon profile. Since we are defining $r_{\rm ap}$ as some fraction of each galaxy's physical scale ($R_e$), we multiply $r_{\rm ap}$ by the minor-to-major axis ratio ($b/a$) so that $r_{\rm ap}$ covers the defined radius in projection with the face of the galaxy when placed along, but perpendicular to, the expected major axis. This scaling by $b/a$ is done when calculating $R_{\rm AB}$ for all subsequent models as well as real data.

Larger values of $r_{\rm ap}$ will tend to yield less noisy measurements of $R_{\rm AB}$ by being less sensitive to random errors and small-scale variations in a velocity field (e.g., turbulent motions) compared to smaller values of $r_{\rm ap}$.  At the same time, as $r_{\rm ap}$ gets larger, true PA$_k$ variations tend to get smoothed out.  This is seen in Fig.~\ref{fig:Radon_model_vary_aperture}, where larger values of $r_{\rm ap}$ fail to capture the true PA$_k$ at $\rho\sim0$, making the overall range of $\thetahat$ smaller, although the qualitative behavior (whether $\thetahat$ is constant or varying with radius) is still apparent. Additionally, as $r_{\rm ap}$ increases, $R_{\rm AB}$ is more susceptible missing data. When $r_{\rm ap}=2R_e$, $\thetahat$ can only be reliably estimated near $\rho=0$ because else the Radon aperture extends beyond the edge of the disk with detectable emission (this specific example is dependent on the chosen disk size of our model, but the point remains). Also apparent in Fig.~\ref{fig:Radon_model_vary_aperture} are sudden changes in $\thetahat$ at large $\rho$. These features are caused by missing data at large $\rho$, which biases our estimate of $\thetahat$ towards values where it can be measured.  The final step of our tracing algorithm from Section~\ref{sec:tracing_algorithm} helps flag and remove these value from any analysis, but we leave them in Fig.~\ref{fig:Radon_model_vary_aperture} so the reader is made aware of the existence of this bias.

Taking into account the pros and cons of different choices of $r_{\rm ap}$, in Section~\ref{sec:application} we discuss our choice for $r_{\rm ap}$ when applying our algorithm to the MaNGA data set.

\begin{figure*}
	\includegraphics[width=2\columnwidth]{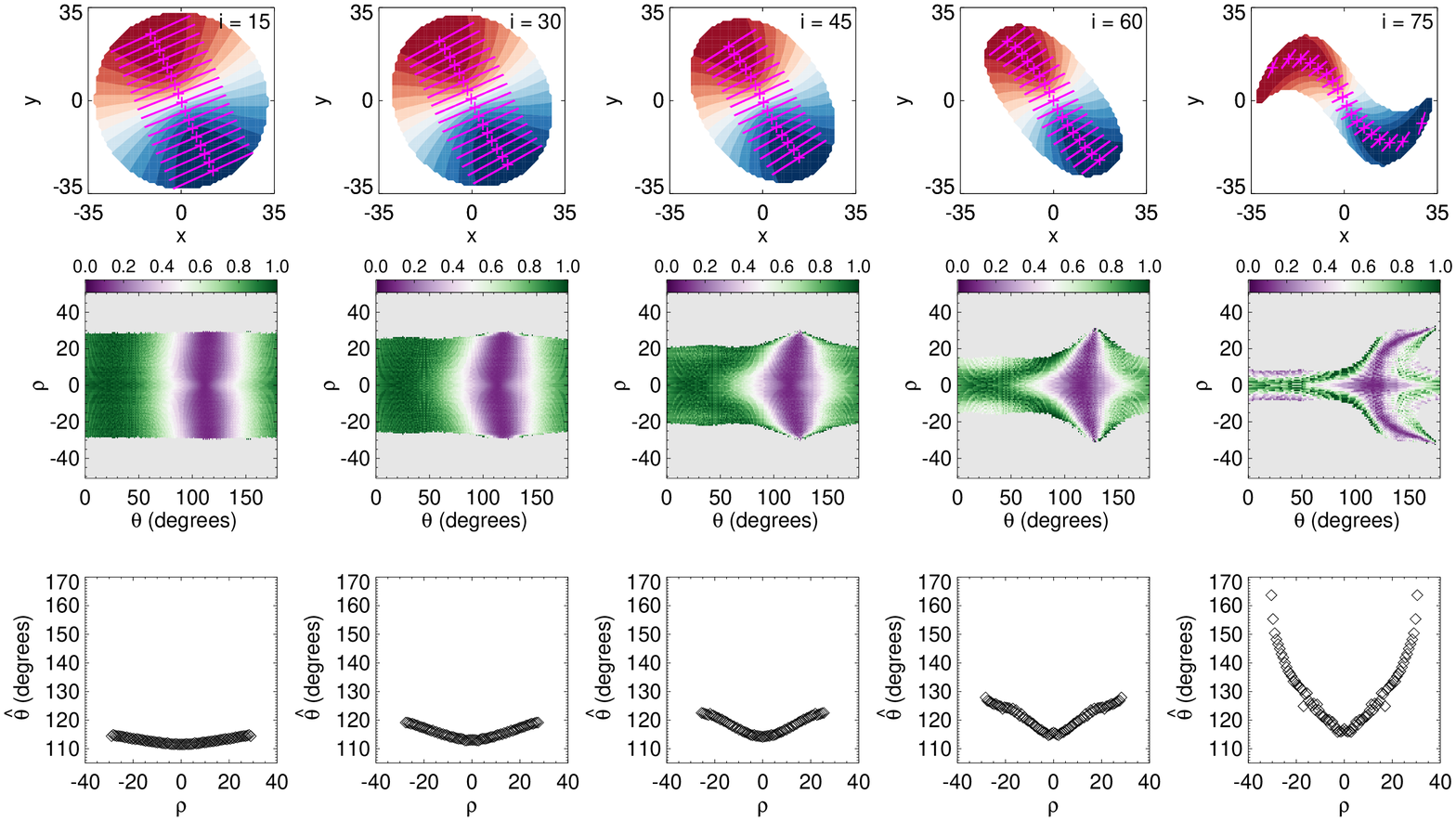}
    \caption{Same as Fig.~\ref{fig:Radon_model_vary_aperture} but showing variations in $R_{\rm AB}$ and $\hat{\theta}(\rho)$ as a function of central inclination with respect to the sky plane (indicated to the upper right of each velocity field) for a model galaxy whose intrinsic inclination relative to the center is changing constantly at a rate of $10^{\circ}\,R_e^{-1}$.}
    \label{fig:Radon_model_vary_inclin}
\end{figure*} 

\subsubsection{Center Definition}
\label{sec:sys_err_center}

By default, our algorithm assumes that the kinematic center of a velocity field lies at the very center of the input $v(x,y)$ grid.  In reality, this is not always the case. For instance, MaNGA IFUs are often positioned on the photometric center of a galaxy based on SDSS imaging, but in some cases they are purposefully repositioned (although this is typically done to correct a poor previously determined photometric center; \citealt{Wake17}).  However, kinematic centers do not necessarily coincide with photometric centers of galaxies (e.g., \citealt{GarciaLorenzo15}), and  particularly for low surface brightness dwarf galaxies, photometric centers can be poorly defined.  

In Fig.~\ref{fig:Radon_model_vary_offcenter}, we demonstrate the impact of incorrect centers on $R_{\rm AB}$ and the derived Radon profile for a model galaxy with a constant PA$_k$. For this model, being off center by more than a few pixels can introduce differences of $\sim10^{\circ}$ or more between the two sides of the Radon profile.  In Section~\ref{sec:recentering_method}, we discuss a method for determining the optimum center of a velocity field using the asymmetry in $R_{\rm AB}$ as a guide.

\subsubsection{Inclination}
\label{sec:sys_err_inclination}

Inclination-dependent projection effects can hide distortions in a velocity field.  To illustrate this issue, Fig.~\ref{fig:Radon_model_vary_inclin} shows $R_{\rm AB}$ and Radon profiles for a model velocity field viewed at varying inclinations (indicated in the upper-right corner of each panel in the top row). In this example, we create a more realistic warped disk model where the {\it intrinsic} galaxy inclination relative to the center changes at a constant rate of $10^{\circ}\,R_e^{-1}$.  The measured change in $\thetahat(\rho)$ caused by the warp in the disk become substantially weaker as the galaxy becomes more face-on, although the qualitative behavior of $\thetahat(\rho)$ is always present.  Nevertheless, the absolute strength of the change in inclination which drives the variation in PA$_k$ cannot be determined without additional information.

Note that we have only illustrated one particular type of distortion, a warped disk, where the distortions are most apparent in the edge-on case. Alternatively, distortions may be driven by features in the plane of the disk, such as bars or spiral arms, which may become difficult to detect at extremely high inclinations.

\section{Application to MaNGA velocity fields}
\label{sec:application}

We now apply the Radon transform to data from the SDSS-IV MaNGA (Mapping Nearby Galaxies at APO) survey, an integral field unit (IFU)
survey of 10,000 $z{\sim}0$ galaxies with stellar masses $M_*\gtrsim 10^9\,\msun$ \citep{Bundy15,Drory15,Law15,Yan16b,Yan16a,Law16,Blanton17}. MaNGA uses the SDSS 2.5m telescope \citep{Gunn06} and BOSS
spectrographs \citep{Smee13}, with a wavelength coverage of \mbox{3500--10000 {\AA}}, spectral resolution $R\sim2000$
(instrumental resolution $\sigma{\sim}60\,{\rm km\,s^{-1}}$), and an effective spatial resolution of 2.5$\arcsec$ (FWHM) after combining dithered observations. For this work, the parent sample is composed of the 2776 galaxies released as part of SDSS DR14 \citep{Abolfathi18}. We use galaxies from both the Primary and Secondary MaNGA samples which have radial coverage out to 1.5$R_e$ and 2.5$R_e$, respectively \citep{Wake17}.

\subsection{Velocity Extraction}

Gas and stellar velocity fields come from the internal MaNGA Product Launch 5 (MPL-5) of the MaNGA Data Analysis Pipeline (DAP). The DAP products used in this work differ slightly from those that will be released publicly as part of SDSS DR15 (MPL-7). A full description of DAP will be presented in Westfall et al. (in preparation), but we briefly summarize the procedure here.

Starting with the output from the Data Reduction Pipeline (DRP; \citealt{Law16}), which provides fully reduced, background subtracted data cubes with $0.5\arcsec\times0.5\arcsec$ spaxels, the penalized pixel fitting algorithm (pPXF; \citealt{Cappellari04}) is applied to each binned spectrum. This algorithm fits a linear combination of template galaxy spectra convolved to a line of sight velocity distribution.  Any regions of the spectrum flagged as unreliable by the DRP, or with known emission lines, are ignored.  Once the best fitting stellar continuum model is determined, it is subtracted from each spectrum, and each emission line is fit separately with a Gaussian profile.  We use the fits to the H$\alpha$ emission line as our indicator of gas velocity.

Before calculating $R_{\rm AB}$, we apply a few additional quality cuts.  First, we remove any spaxels flagged by the DRP or DAP. We then remove spaxels with low signal-to-noise (S/N) ratio, either ${\rm S/N} < 3$ on the H$\alpha$ flux measurement from a Gaussian fit or ${\rm S/N} < 3$ in the continuum flux, for the gas and stellar velocity fields, respectively.  We then check for any remaining highly deviant velocity measurements by comparing the velocity in each spaxel with the median of all velocities within a 5x5 box around that spaxel, and remove it if the value differs from the median by more than $\Delta V/2$, where $\Delta V$ is the absolute difference between the two velocities which enclose 95\% of all measured velocities for that galaxy.

\subsection{Practical Calculation of $R_{\rm AB}$}

When calculating $R_{\rm AB}(\theta,\rho)$, we estimate the velocities along each line segment using nearest neighbor interpolation. Our calculation follows the formalism implemented by Interactive Data Language (IDL)\footnote{\url{http://www.harrisgeospatial.com/docs/Radon.html}}, where the Radon transform equation is first rotated by $\theta$, and the transformation is then broken into two regimes, one for $\theta \leq 45^{\circ}$ and $135^{\circ} \leq \theta \leq 180^{\circ}$, and one for $45^{\circ} \geq 135^{\circ}$, i.e. shallower and steeper lines, respectively.  The new transformations are written as:
\begin{equation}
R(\theta,\rho) = 
\begin{cases}
\frac{\Delta x}{\vert \sin{\theta}\vert}\sum_x v(x_i,[a_1x_i+b_1])-\tilde{v} & \vert \sin{\theta}\vert > \frac{\sqrt{2}}{2} \\
\frac{\Delta y}{\vert \cos{\theta}\vert} \sum_y v([a_2y_i+b_2],y_i)-\tilde{v} & \vert \sin{\theta}\vert \leq \frac{\sqrt{2}}{2} \\
\end{cases}
\end{equation}
where $\Delta x$ and $\Delta y$ are the sample steps in the $x$ and $y$ directions (1 spaxel in our case), and $\tilde{v}$ is the median of all velocity measurements within $\pm r_{\rm ap}$ of [$\theta_i,\rho_i$]. The square brackets indicate rounding to the nearest integer value. The slopes and intercepts in the above transformation are given by
 \begin{eqnarray}
 a_1 = \frac{\Delta x\cos{\theta}}{\Delta y \sin{\theta}} \\
 b_1 = \frac{\rho - x_{\rm min}\cos{\theta}-y_{\rm min}\sin{\theta}}{\Delta y \sin\theta} \\
 a_2 = \frac{1}{a_1} \\
 b_2 = b_1\frac{\sin{\theta}}{\cos{\theta}}
 \end{eqnarray}
Our own custom IDL program used to calculate $R$, $R_{\rm A}$, and $R_{\rm AB}$ throughout this work is available online.\footnote{\url{https://github.com/dvstark/radon-transform}}

\subsection{Choice of $r_{\rm ap}$}
\label{sec:rap_choice}

As discussed in Section~\ref{sec:sys_apsize}, $r_{\rm ap}$ must be chosen to strike a balance between being small enough to be both sensitive to variations in the PA$_k$ and minimally affected by proximity to the edge of the velocity field, while large enough not to be significantly affected by noise or turbulent motions. To enable a consistent analysis of all galaxies, we also want $r_{\rm ap}$ to be the same size relative to some characteristic size scale of each galaxy, such that $r_{\rm ap}=\alpha R_e$, where $\alpha$ is a constant and $R_e$ is the half-light radius. Furthermore, as discussed above when calculating $R_{\rm AB}$ for model velocity fields, we also want to ensure $r_{\rm ap}$ scales as $\cos{i}$ so that  the integrals in $R_{\rm AB}$ are calculated over approximately the same relative physical scale projected onto the face of each galaxy disk. In practice, we assume $\cos{i}\sim b/a$ where $b/a$ is the minor to major axis ratio, such that $r_{\rm ap}=\alpha R_e\times b/a$. For our analysis, $R_e$ and $b/a$ are the elliptical Petrosian half-light radius and minor-to-major axis ratio taken from the NASA Sloan Atlas (NSA)\footnote{\url{http://www.nsatlas.org/}}

Based on Fig.~\ref{fig:Radon_model_vary_aperture}, setting $\alpha \leq 1/2$ does the best job of tracking the true PA$_k$ to the largest possible radius. However, this choice of $r_{\rm ap}$ means that $R_{\rm AB}$ will calculated over spatial scales that are smaller than the MaNGA spatial resolution (typically ${\sim}2.5\arcsec$; \citealt{Law16}) for 25\% of galaxies.  As a compromise, we set $\alpha=1$ (corresponding to the 3rd case in Fig.~\ref{fig:Radon_model_vary_aperture}), which ensures $R_{\rm AB}$ is calculated over spatial scales larger than the typical spatial resolution for 99\% of galaxies, albeit with the risk that some radial variation in PA$_k$ may be blurred out.

\subsection{Recentering Method}
\label{sec:recentering_method}

Errors in the assumed kinematic center can induce artificial variations in Radon profiles (see Section~\ref{sec:sys_err_center}). To mitigate this issue, we use $R_{\rm AB}$ itself to find the best kinematic center under the assumption that the true center is that where the asymmetry in $R_{\rm AB}$ is minimized. Similar approaches have been adopted when calculating asymmetries in imaging data and rotation curves \citep{Conselice00, Kannappan02}. 

Our recentering procedure is as follows.  For each velocity field, we define a $7\times7$ grid with a spacing of 0.25 spaxels centered on the photometric center of the galaxy.  We shift the velocity field center around this grid, using bilinear interpolation to estimate the velocity field each time.  At each grid position, we recalculate $R_{\rm AB}$, trace $\hat{\theta}(\rho)$ following the procedure in Section~\ref{sec:tracing_algorithm}, and calculate the asymmetry as:
\begin{equation}
\label{eq:asym_rc}
A_{i,j} = \frac{\sum{\left| \hat{\theta} - \hat{\theta}_{\rm flip} \right|}}{2N_{i,j}}w_{i,j}
\end{equation}
where $\hat{\theta}_{\rm flip}$ is the reversed $\hat{\theta}$ array, and $N_{i,j}$ is the number of values in the $\hat{\theta}$ array when calculated at the currently adopted center. $w$ is a weight factor defined as
\begin{equation}
w_{i,j}=\frac{N_{0,0}}{N_{i,j}}
\end{equation}
where $N_{0,0}$ is the number of values in the $\hat{\theta}$ array when using the original photometric center.  This weight factor helps account for differences in the asymmetry that can arise when there are a different number of individual $\hat{\theta}$ measurements at a given adopted center, which essentially artificially raises/decreases the measured asymmetry for regions where the $\hat{\theta}$ arrays are smaller/larger. The best center is taken to be the point where $A$ is minimized.  If the derived center lies on the edge of the 7x7 grid, we expand the grid by a factor of two and rerun the algorithm.  If the best determined center is still at the edge of the grid, we do not expand it further because at this point the center is extremely far from the photometric center and likely not well-determined. These galaxies are rejected from the analysis, but this issue only occurs in $\sim$1-2\% of cases. This entire procedure is done independently for gas and stellar velocity fields. The average difference between the IFU and estimated velocity field centers is 1$\arcsec$. The magnitude and direction of the positional shifts of the gas and stellar velocity fields are only weakly correlated but at a statistically significant level (Spearman rank correlation coefficient of 0.1 with a 0.06\% chance of the correlation occurring randomly). 

\subsection{Uncertainty Estimation}
\label{sec:uncertainty}

As discussed in \citet{Law16}, the creation of the rectilinearly sampled spectral data cubes using flux measurements from individual fibers leads to significant covariance between adjacent spaxels.  We find that ignoring such covariances will significantly underestimate uncertainties on $R_{\rm AB}$ and $\thetahat$, so we have taken steps to account for covariance in each stage of our analysis.

We first assume the correlation matrix of velocity field spaxels is approximately
\begin{equation}
c_{i,j} = 
\begin{cases}
e^{-0.5\left(\frac{d_{i,j}}{1.9}\right)^2} & d_{i,j} < 6.4 \\
0 & d_{i,j}>6.4
\end{cases}
\end{equation}
where $d_{i,j}$ is the distance between two spaxels in units of pixels (Westfall et al., in preparation). This  $N\times N$ correlation matrix combined with the estimated velocity errors yields the full covariance matrix for the velocity field, $C_v$. For $R_{\rm AB}$ with $M$ pixels, its  $M\times M$ covariance matrix is calculated as 
\begin{equation}
C_R=W\times C_v \times W^T
\end{equation}
where $W$ is an MxN matrix of 1 or 0 indicating which velocity spaxels are included in the calculation $R_{\rm AB}$ at each position. 

To estimate the uncertainty on $\thetahat$, we opted for a simple Monte-Carlo approach where for each $\rho$ in $R_{AB}$, we repeat the fitting step described in Section~\ref{sec:tracing_algorithm} 100 times, each time adding random noise to the data using the full covariance matrix $C_R$ (not just the diagonal elements). Note that adding random noise at this stage requires $C_R$ be invertible, which we found was not always the case due to numerical rounding errors that made $C_R$ non-positive definite.  However, applying a small (typically 1\%) scale factor to the diagonal elements of $C_R$ solved this problem, and such a small offset has little impact on the uncertainties propagated through our analysis. The median and standard deviation of the 100 fitted centroids are taken as the final estimate of $\thetahat$ and its uncertainty. 

We have tested the impact that ignoring covariance has on our final analysis.  Assuming all errors are independent can result in underestimating the uncertainty on $\thetahat$ by a factor of $\sim$5 on average, but with a large tail towards higher values.

\begin{figure*}
	\includegraphics[width=1.95\columnwidth]{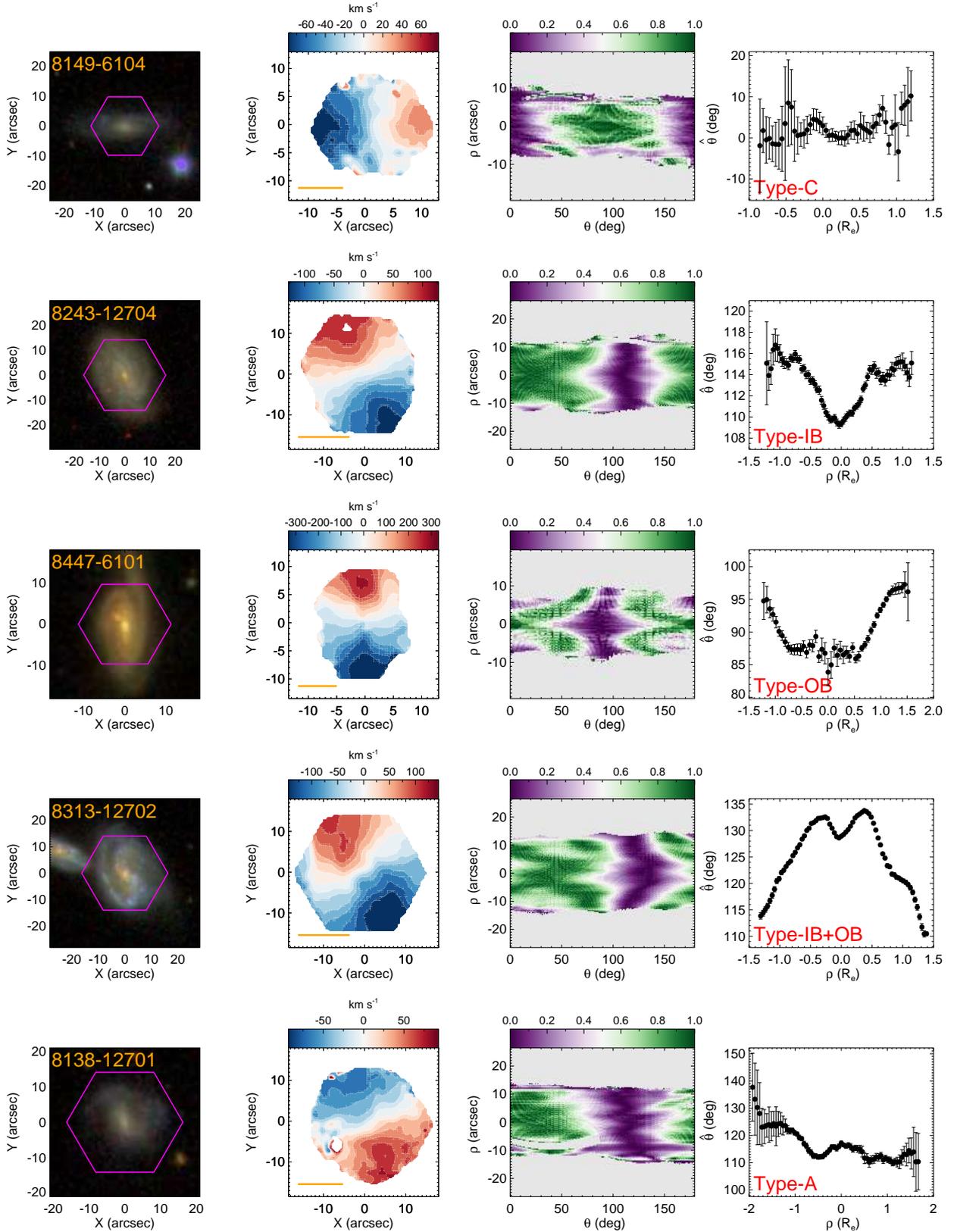}
    \caption{Example output from the Radon transform applied to MaNGA H$\alpha$ velocity fields.  From left to right, the panels show: (i) An SDSS $gri$ cutout of the galaxy with the hexagonal MaNGA IFU bundle shape overlaid in magenta. The number in the upper right corner indicates the PLATE-IFU designation of the observation. (ii) The H$\alpha$ velocity field with the size of the radon aperture indicated by the orange line in the lower-left corner. (iii) The resulting rescaled map of $R_{\rm AB}$. (iv) The derived Radon profile and uncertainty. The Radon profile classification is noted in the bottom-left corner. }
    \label{fig:Radon_example_output}
\end{figure*}

\subsection{Characteristic Output}
\label{sec:characteristic_output}

\begin{figure}
\includegraphics[width=\columnwidth]{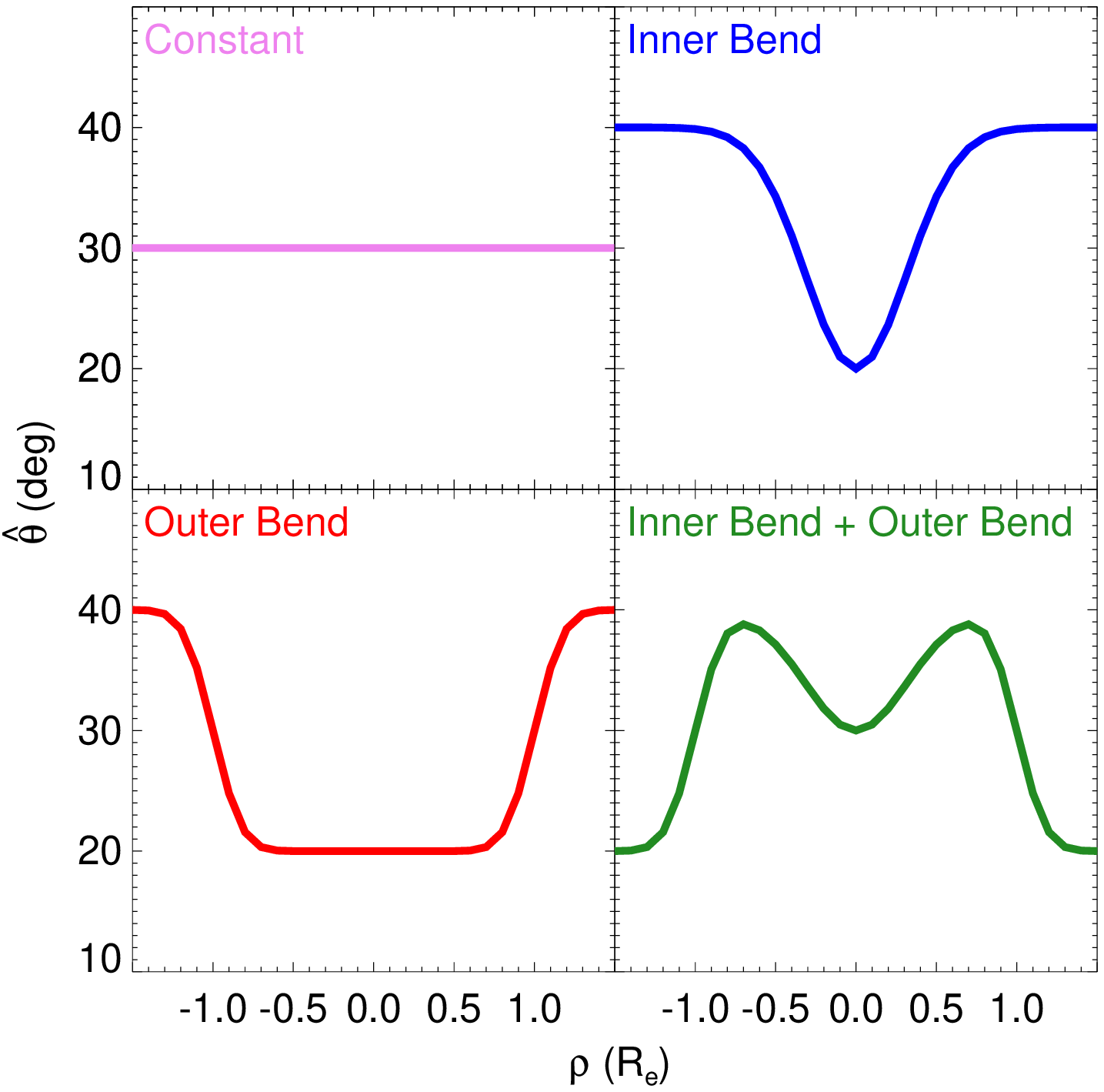}
\caption{Examples of models representing the first four Radon profile classifications.}
\label{fig:fit_models}
\end{figure}
In Fig.~\ref{fig:Radon_example_output} we show example output maps of $R_{\rm AB}$ and $\thetahat(\rho)$ for MaNGA H$\alpha$ velocity fields. Each of the examples represents a commonly occurring pattern in the data.  Based on these patterns, we divide Radon profiles into five major classes: 
\begin{description}
\item {\bf 1:} Radon profiles that are consistent with having a fixed $\thetahat$ at all radii.  We refer to these profiles as {\bf Constant}, or {\bf Type-C}. \vspace{5pt}
\item {\bf 2:} Radon profiles where symmetric variations in $\thetahat$ begin immediately at $\vert \rho \vert >0$, typically settling to a constant value by $0.5-1R_e$ (see also Section~\ref{sec:structure}). These profiles are modeled by a Gaussian function
\begin{equation}
\label{eq:gaussian}
\hat{\theta}(\rho) = Ae^{\frac{-\rho^2}{2B^2}}+C \vspace{5pt}
\end{equation}
We refer to these cases as {\bf Inner Bends}, or {\bf Type-IB}.
\item {\bf 3:} Galaxies with constant $\hat{\theta}$ at small radius which then transitions to a different $\hat{\theta}$ at some larger radius. These profiles are well-described by a Busy Function \citep{Westmeier14}:
\begin{equation}
\label{eq:busy}
\hat{\theta}(\rho) = \frac{A}{4}(\erf(B(W+\rho))+1)(\erf(B(W-\rho))+1)+C
\end{equation}
We refer to these profiles as {\bf Outer Bends}, or {\bf Type-OB}.
\item {\bf 4:} Radon profiles which show properties of both Type-IB and Type-OB profiles. These are modeled using a combination of the Gaussian and Busy functions
\begin{equation}
\hat{\theta}(\rho) = \frac{A}{4}(\erf(B(W+\rho))+1)(\erf(B(W-\rho))+1)Ce^{D\rho^2}+E
\end{equation}
We refer to these as {\bf Inner Bend + Outer Bend}, or {\bf Type-IB+OB}.
\item {\bf 5:} Galaxies with {\bf asymmetric} Radon profiles, or {\bf Type-A}. We estimate asymmetry with two parameters:
\begin{equation}
\label{eq:a2}
A_1=\frac{1}{\Delta{\hat{\theta}}}\frac{\sum_i{w_i\delta\hat{\theta}}}{\sum_i{w_i}}
\end{equation}
\begin{equation}
\label{eq:a1}
A_2=\sum_i{\frac{\delta\hat{\theta}}{\sigma_{\delta\hat{\theta},i}}}
\end{equation}
where $\delta\thetahat_i$ is the absolute magnitude of the difference between $\thetahat_i$ and the value on the opposite side of the Radon profile (i.e., same $\vert\rho\vert$ but opposite sign), $\sigma_{\delta\thetahat}$ is the corresponding uncertainty on $\delta\thetahat$, $w_i$ is a weight term defined as $w_i=\sigma_{\delta\thetahat}^{-2}$, and $\Delta\thetahat$ is the range of $\thetahat$ that encloses 95\% of the measured values.  $A_1$ indicates a fractional asymmetry relative to the overall variation $\thetahat$, and is very similar to the definition of asymmetry defined in Eq.~\ref{eq:asym_rc} except that the data points are weighted. $A_2$ indicates whether two sides of a Radon profile are significantly different relative to their uncertainties. Our final asymmetry category is defined as anything with $A_1>0.2$ and $A_2>3$, i.e., the two sides must differ by a significant fraction of the overall range in $\thetahat$ and be unexplainable by measurement uncertainty.
\end{description}

These five categories are meant to be phenomenological, and were initially created based on the observed patterns seen in Radon profiles without any additional information. However, it is fair to say that certain models may be well-suited to capture certain physical processes. For example, the Busy function used to represent Type-OB profiles capture behavior we might expect from galaxies with misaligned rotation in the outer disk, while Type-IB profiles may be a better representation of phenomena at smaller radii, such as bar distortions.

\subsection{Automated Radon Profile Classification}
\label{sec:classification}

We assign each galaxy's Radon profile into one of the five categories described in the previous section. The classification of Type-A profiles is straightforward and is simply based on the values estimated with Eqs.~\ref{eq:a1} and \ref{eq:a2}.  For the first four categories, we employ an automated scheme whereby we conduct a maximum likelihood fit of each model to the measured Radon profile, and then use the  Bayesian Information Criterion \citep[BIC;][]{Schwartz78} to determine which model presents the best description of the data without overfitting. The BIC is defined as
\begin{equation}
BIC = p\ln{n}-2\ln\hat{L}
\end{equation}
where $n$ is the number of data points, $p$ is the number of free parameters in the model, and $\hat{L}$ is the maximum likelihood of the model. When comparing two models, $M_1$ and $M_2$, where $M_2$ has more free parameters, we take \mbox{$\Delta BIC(M_2\vert M_1) = BIC(M_2)-BIC(M_1) > 2$} as significant evidence that $M_2$ is favored over $M_1$ \citep{Kass95}. When comparing different models, if there is more than one more complex model with $\Delta BIC>2$, we first choose the simplest of these alternative models, but then estimate the $\Delta BIC$ between this new model and the remaining more complex models.  As an example, if we calculate $\Delta BIC(IB\vert C) = 4.2$ and $\Delta BIC(OB|C) = 5$, we will prefer the Type-IB model.  However, we will then calculate $\Delta BIC(OB|IB)$, and if it is $<2$, the Type-IB model will still be preferred.  The same calculation is done when analyzing whether the Type-IB+OB model is a better description of the data.

As in earlier steps of our analysis, the maximum likelihood estimates should take into account the covariance between data points. The full covariance matrices of our final Radon profiles are unknown, but given the structure of our velocity field covariance matrices, we assume the Radon profile covariance behaves in such a way that the correlation between data points declines with increasing distance between them.  We use the \texttt{george} Python package \citep{Ambikasaran15} to conduct Gaussian process regression and find the maximum likelihood fits of our four models described in Section~\ref{sec:characteristic_output} , while simultaneously fitting a model covariance matrix of the form:
\begin{equation}
C_{ij}=a\exp{\left(\frac{-d_{ij}^2}{2b^2}\right)}
\end{equation}
where $d_{ij}$ is the distance between the ith and jth data points, and $a$ and $b$ are free parameters. Building the covariance into the fitting model typically yields maximum likelihoods that are more conservative than if we were to ignore covariance.

Although we attempted to refine the estimates of the centers of each galaxy before calculating $R_{\rm AB}$, we occasionally find that allowing a slight shift in the definition of the center when fitting the models can significantly improve the resulting fits. This tendency is likely a result of the relatively course grid in our recentering algorithm (Section~\ref{sec:recentering_method}). Therefore, we fit each Radon profile twice, first fixing the center at $\rho=0$ and then allowing the center to vary within $\pm R_e/4$.  We again use the BIC as described above to compare the best fitting model with the center fixed versus with the center allowed to vary, and take whichever model is favored as the final best description of the data.  

\section{A Census of Kinematic Behavior in MaNGA}
\label{sec:census}

Using the automated classification approach described in Section~\ref{sec:classification}, we assign all velocity fields into one of the five types described in Section~\ref{sec:characteristic_output}. In Table~\ref{table:warp_class_fractions} we give the fraction of galaxies that fall into each category among those that were classified.  We have made no cut on inclination or $b/a$, but remind the reader that certain distortions may be missed at extreme inclinations. Unless otherwise stated, we limit our census to galaxies where we can measure $\thetahat$ out to at least $\vert\rho\vert=R_e$, but the model fitting itself extends to as far out as the data is measured. This cut results in a final sample of 907 gas Radon profiles and 936 stellar Radon profiles, among which 466 galaxies have both stellar and gas Radon profiles. The requirement that Radon profiles extend to at least $\vert R_e \vert $ means our analysis of gas velocity fields does not include many red, weakly star forming galaxies which tend to be gas poor and have weak H$\alpha$ emission.  Similarly, our analysis of stellar velocity fields does not include many star-forming dwarf galaxies which tend to have low continuum surface brightness (see also Section~\ref{sec:color_mass}). 

Table~\ref{table:warp_class_fractions} also provides the fraction of galaxies in each classification separately for the primary and secondary MaNGA samples which have radial coverage out to 1.5 and 2.5$R_e$, respectively.  The percentages do not typically vary significantly between the primary, secondary, or combined samples.  We also provide the percentages in each category for gas and stars independent of whether the other component passed our sample selection, as well as for a subsample where both gas and stellar velocity fields for each galaxy had classifications.  Analyzing the behavior of the gas and stars separately, versus analyzing the gas and stars in a sample where both velocity fields have been characterized, does not significantly change the percentages which fall into each category.

Key findings from Table~\ref{table:warp_class_fractions} are:
\begin{itemize}
\item Type-C profiles are most common for both gas and stars, although they are roughly 1.5 times more common in stellar velocity fields.  Although it is the most common classification, the majority of galaxies do not have Type-C gas radon profiles.
\item Type-A profiles are roughly twice as common among gas velocity fields compared to stellar velocity fields.  
\item Inner Bend and Outer Bend profiles occur at similar frequencies in stellar and gas velocity fields. 
\item Type-IB+OB profiles are roughly three times more common in gas velocity fields than stellar velocity fields. In stellar velocity fields, Type-IB+OB profiles comprise only a few percent of the whole population.
\end{itemize}

\begin{table*}
\caption{Percentage of galaxies in each $\hat{\theta}(r)$ category}
\begin{threeparttable}
\begin{tabular}{l|c|c|c|c|c|}
& Constant & Inner Bend & Outer Bend & Inner Bend + Outer Bend & Asymmetric\\
& (Type-C) & (Type-IB) & (Type-OB) & (Type-IB+OB) & (Type-A) \\
\hhline{======}\\
\multicolumn{6}{c}{Gas (All available Radon profiles)} \\
\hline
Primary + Secondary & $32.5_{- 1.5}^{+ 1.6}$ &  $23.6_{- 1.4}^{+ 1.4}$ &  $11.8_{- 1.0}^{+ 1.1}$ &  $ 9.7_{- 0.9}^{+ 1.0}$ &  $22.4_{- 1.3}^{+ 1.4}$ \\ 
Primary & $29.3_{- 2.1}^{+ 2.2}$ &  $25.5_{- 2.0}^{+ 2.1}$ &  $11.1_{- 1.4}^{+ 1.6}$ &  $ 9.8_{- 1.3}^{+ 1.5}$ &  $24.4_{- 2.0}^{+ 2.1}$ \\ 
Secondary & $35.7_{- 2.2}^{+ 2.3}$ &  $21.7_{- 1.9}^{+ 2.0}$ &  $12.5_{- 1.5}^{+ 1.6}$ &  $ 9.6_{- 1.3}^{+ 1.5}$ &  $20.4_{- 1.8}^{+ 1.9}$ \\
\hline
\multicolumn{6}{c}{Gas (Only those which also have stellar Radon profiles)} \\
\hline
Primary + Secondary & $28.3_{- 2.0}^{+ 2.1}$ &  $27.0_{- 2.0}^{+ 2.1}$ &  $10.7_{- 1.3}^{+ 1.5}$ &  $10.7_{- 1.3}^{+ 1.5}$ &  $23.2_{- 1.9}^{+ 2.0}$ \\ 
Primary & $24.8_{- 2.7}^{+ 2.9}$ &  $28.5_{- 2.8}^{+ 3.0}$ &  $11.2_{- 1.9}^{+ 2.2}$ &  $ 9.5_{- 1.7}^{+ 2.0}$ &  $26.0_{- 2.7}^{+ 2.9}$ \\ 
Secondary & $32.1_{- 3.0}^{+ 3.2}$ &  $25.4_{- 2.8}^{+ 3.0}$ &  $10.3_{- 1.8}^{+ 2.2}$ &  $12.1_{- 2.0}^{+ 2.3}$ &  $20.1_{- 2.5}^{+ 2.8}$ \\ 
\hline
\multicolumn{6}{c}{Stars (All available Radon profiles)} \\
\hline
Primary + Secondary & $54.1_{- 1.6}^{+ 1.6}$ &  $20.6_{- 1.3}^{+ 1.3}$ &  $ 9.0_{- 0.9}^{+ 1.0}$ &  $ 2.7_{- 0.5}^{+ 0.6}$ &  $13.7_{- 1.1}^{+ 1.2}$ \\ 
Primary & $53.0_{- 2.3}^{+ 2.3}$ &  $18.3_{- 1.7}^{+ 1.8}$ &  $11.0_{- 1.3}^{+ 1.5}$ &  $ 2.7_{- 0.6}^{+ 0.8}$ &  $15.0_{- 1.5}^{+ 1.7}$ \\ 
Secondary & $55.2_{- 2.3}^{+ 2.3}$ &  $23.1_{- 1.9}^{+ 2.0}$ &  $ 6.8_{- 1.1}^{+ 1.3}$ &  $ 2.6_{- 0.7}^{+ 0.9}$ &  $12.3_{- 1.5}^{+ 1.6}$ \\ 
\hline
\multicolumn{6}{c}{Stars (Only those which also have gas Radon profiles)} \\
\hline
Primary + Secondary & $54.9_{- 2.3}^{+ 2.3}$ &  $22.7_{- 1.9}^{+ 2.0}$ &  $ 9.0_{- 1.2}^{+ 1.4}$ &  $ 1.7_{- 0.5}^{+ 0.7}$ &  $11.6_{- 1.4}^{+ 1.6}$ \\ 
Primary & $53.3_{- 3.2}^{+ 3.2}$ &  $19.8_{- 2.4}^{+ 2.7}$ &  $10.3_{- 1.8}^{+ 2.1}$ &  $ 2.1_{- 0.7}^{+ 1.1}$ &  $14.5_{- 2.1}^{+ 2.4}$ \\ 
Secondary & $56.7_{- 3.3}^{+ 3.3}$ &  $25.9_{- 2.8}^{+ 3.0}$ &  $ 7.6_{- 1.6}^{+ 2.0}$ &  $ 1.3_{- 0.6}^{+ 1.0}$ &  $ 8.5_{- 1.7}^{+ 2.0}$ \\ 
\\
\hhline{======}
\end{tabular}
 \begin{tablenotes}
 \item Note -- For each galaxy component, we provide the percentages of galaxies that fall into each classification, both independent of whether the other component has a classification, and specifically for the subset where both components have classifications. The rows labeled ``Primary" and ``Secondary" refer to the primary and secondary MaNGA samples which have radial coverage out to 1.5 and 2.5 $R_e$, respectively \citep{Wake17}. The uncertainties are from binomial statistics and do not reflect any possible classification errors.
 \end{tablenotes}
 \end{threeparttable}
\label{table:warp_class_fractions}
\end{table*}

We also examine whether or not our classifications may be subject to biases with respect to physical resolution (kpc) and/or inclination. Poorer spatial resolution/beam smearing may blur out distortions in velocity fields and bias Radon profiles into certain classifications. The fact that the values presented in Table~\ref{table:warp_class_fractions} do not strongly depend on whether we look at the Primary or Secondary samples (which have different average physical resolution as a function of stellar mass; \citealt{Wake17}) is a good indication that spatial resolution is not affecting our results. We also examined the distributions of physical resolution for each of our Radon profile classifications, but we find no statistically significant difference between any of the groups. Furthermore, if poor spatial resolution is causing Radon profiles to be biased towards certain classifications, it should affect gas and stellar velocity fields equally. However, as we will discuss in Sections~\ref{sec:star_gas_agreement} and \ref{sec:color_mass} the Radon profile classifications for gas and stars are largely independent of one another. 

Additionally, inclination can weaken the magnitude of certain distortions (see Section~\ref{sec:sys_err_inclination}).  Among gas velocity fields, we do find that Type-A profiles have a slight tendency to be found at lower $b/a$ (higher inclination) compared to other classifications (with typical K-S test significances of ${\sim}2.5\sigma$).

\subsection{Structural Properties}
\label{sec:structure}
\begin{figure*}
\includegraphics[width=\columnwidth]{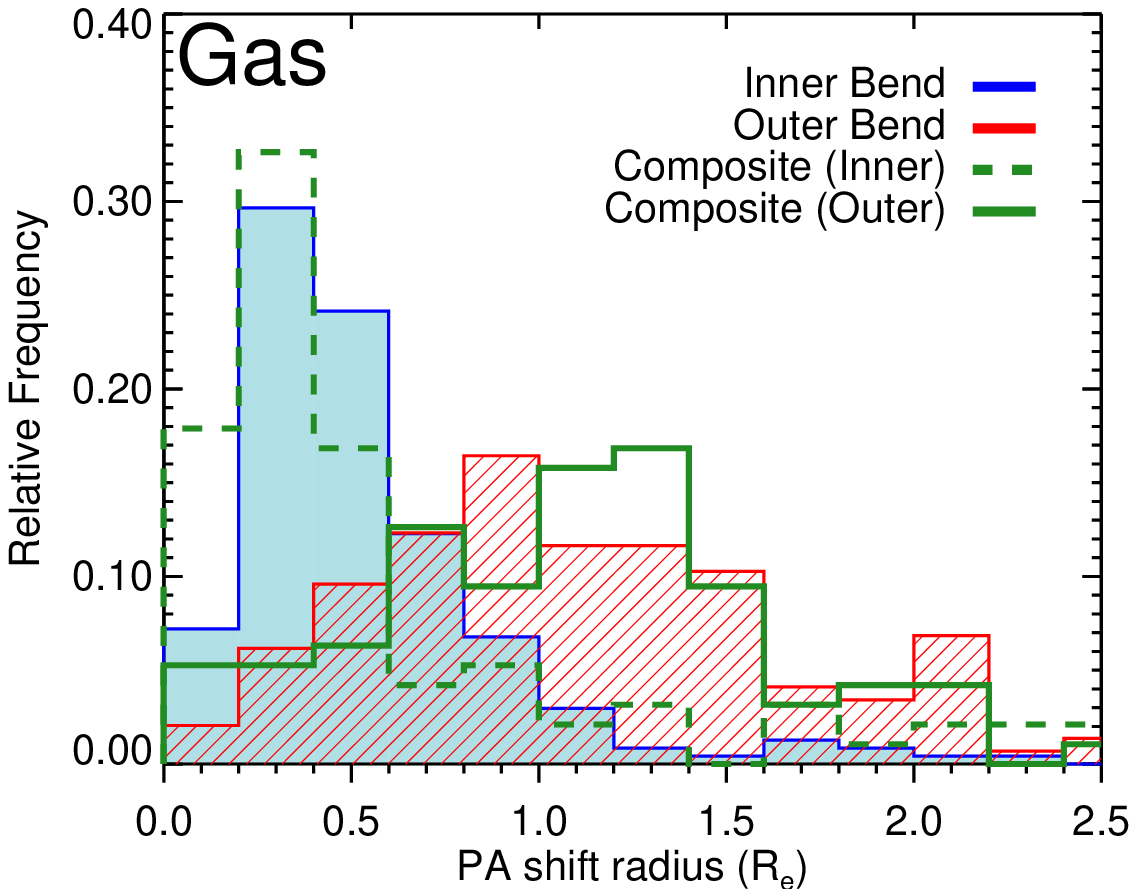}
\includegraphics[width=\columnwidth]{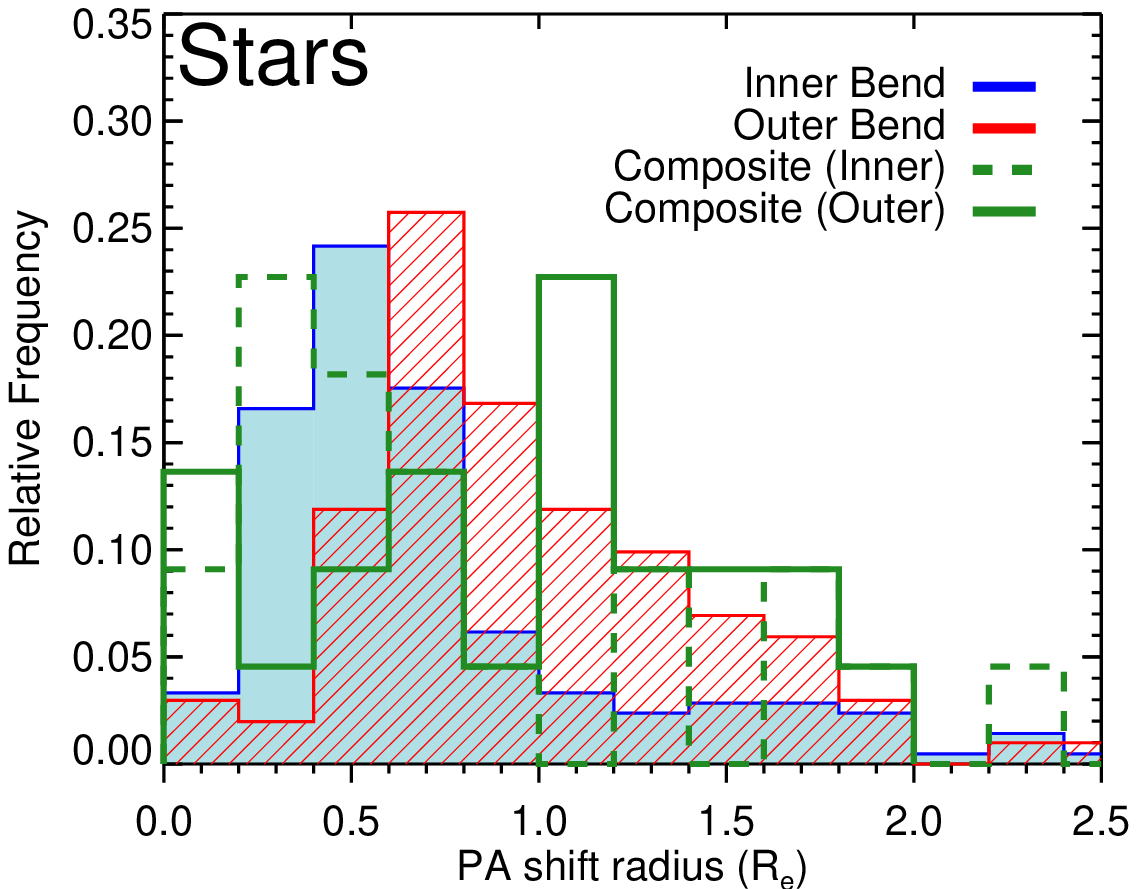}
\caption{Distribution of characteristic radii for Type-IB, Type-OB, and Type-IB+OB Radon profiles of gas and stars, where the characteristic radius is the half-width at half-maximum (HWHM) of the Gaussian or Busy functions used to represent these different classifications. For Type-IB-OB profiles, we separately plot the distribution characteristic radii for the Gaussian and Busy function components (dashed and solid green lines, respectively).}
\label{fig:warp_fit_properties}
\end{figure*}
To further illuminate the differences between different Radon profile classifications, we show the distribution of characteristic radii for Type-IB, Type-OB, and Type-IB+OB profiles in Fig.~\ref{fig:warp_fit_properties}, where this characteristic radius is an indicator of where $\thetahat$ is transitioning to a new value.  For Type-IB profiles, this radius corresponds to the half-width-half-maximum (HWHM) of the Gaussian, or $B\sqrt{2 \ln{2}}$, where $B$ comes from Eq.~\ref{eq:gaussian}. For Type-OB profiles the characteristic radius corresponds to the $W$ parameter in Eq.~\ref{eq:busy}, which again is equal to the HWHM of that function. For Type-IB+OW profiles, we plot the characteristic radii for the Gaussian and Busy components separately (dashed and solid green lines, respectively).  

The distribution of characteristic radii for Type-IB profiles peaks around 0.5--0.6$R_e$ and drops off rapidly at larger radius, while Type-OB profiles have characteristic radii extending to much larger values.  The different distributions for Type-IB and Type-OB profiles is at least partially by design, as the Busy function which represents Type-OB profiles is specifically chosen to be able to capture cases where the shift in PA$_k$ begins away from $\rho=0$. The characteristic radii of the Gaussian and Busy function components of type-IB+OB profiles show similar distributions as the characteristic radii of Type-IB and Type-OB profiles. Notably, there is no strong difference between the distribution of characteristic radii, regardless of type, between gas and stellar velocity fields.

\subsection{Agreement between gas and stars}
\label{sec:star_gas_agreement}

\begin{figure*}
\includegraphics[width=2\columnwidth]{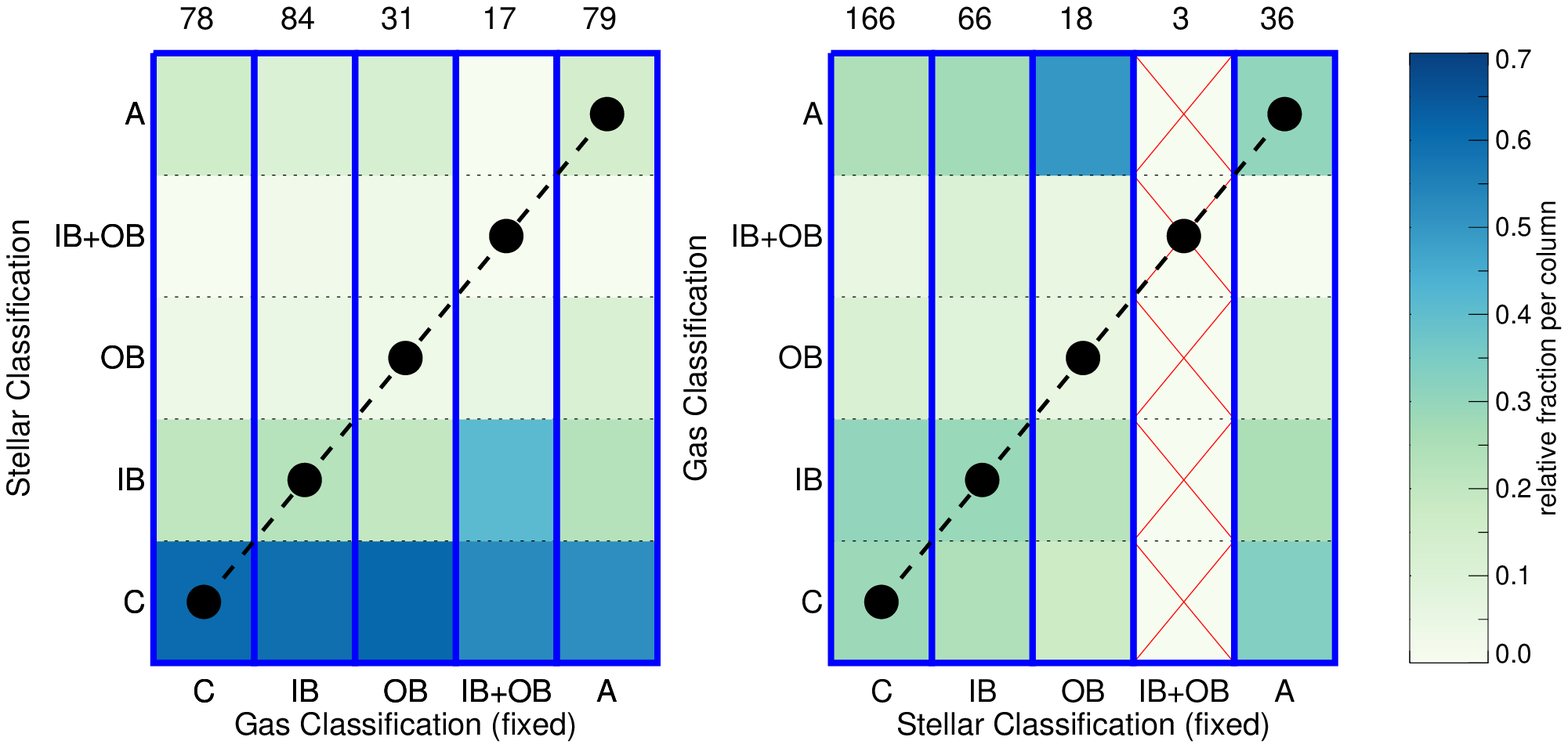}
\caption{{(Left) Relative distribution of stellar Radon profile classifications ($y$-axis) at fixed gas Radon profile classification ($x$-axis). The numbers above each column indicate the total number of galaxies at each fixed gas Radon profile classification, and the colorbar reflects the relative fraction within each column. Black points indicate where the stellar and gas classifications are the same, and if the stellar and gas classifications were in good agreement, the distributions in each column would peak at the black point. (Right) Same as left panel, but showing the relative distribution of gas Radon profile classifications at fixed stellar Radon profile classification. Due to the small number of galaxies with Type-IB+OW stellar Radon profiles, we do not plot their distribution. To ensure a fair comparison, we limit this analysis to a subset of galaxies where $\thetahat$ could be measured to at least $1.5R_e$ for both components, and the classifications only consider the data at $\rho<1.5R_e$.}}
\label{fig:stars_vs_gas_type}
\end{figure*}

Table~1 shows that the fractions of Radon profiles which fall into each of our five classifications is not the same for gas and stars. To further illuminate the extent to which gas and stellar velocity fields behave dissimilarly, {Fig.~\ref{fig:stars_vs_gas_type} shows the distributions of gas and stellar Radon profile classifications in groups of galaxies separated by the other component's classification, e.g., the first column of the left panel shows the distribution of stellar velocity field classifications for the subsample of galaxies for which the gas velocity fields are classified as Type-C. To ensure a fair analysis, we have limited the sample to only galaxies where $\thetahat$ has been measured to at least $1.5R_e$ for both gas and stars, and also limit the model fits to $\vert\rho\vert<1.5R_e$ even if they extend to larger radius.} 

{This figure illustrates that the stellar and gas Radon profile classifications rarely agree with one another (if the classifications were in good agreement, the peak of the distribution in each column of Fig.~\ref{fig:stars_vs_gas_type} would peak at the location of the black point).  The one exception to this statement is when gas velocity fields are classified as Type-C, most (60\%) stellar Radon profiles are as well. The opposite is not true; if stellar velocity fields are classified as Type-C, only ${\sim}30$\% of gas velocity fields are Type-C, the rest mostly being Type-IB or Type-A. Regardless of how a gas Radon profile is classified, the stellar Radon profile will always most likely be classified as Type-C. When the stellar classifications are fixed, the behavior of the gas velocity fields is more diverse.}

\subsection{Color-Stellar Mass Distribution}
\label{sec:color_mass}

\begin{figure}
\includegraphics[width=\columnwidth]{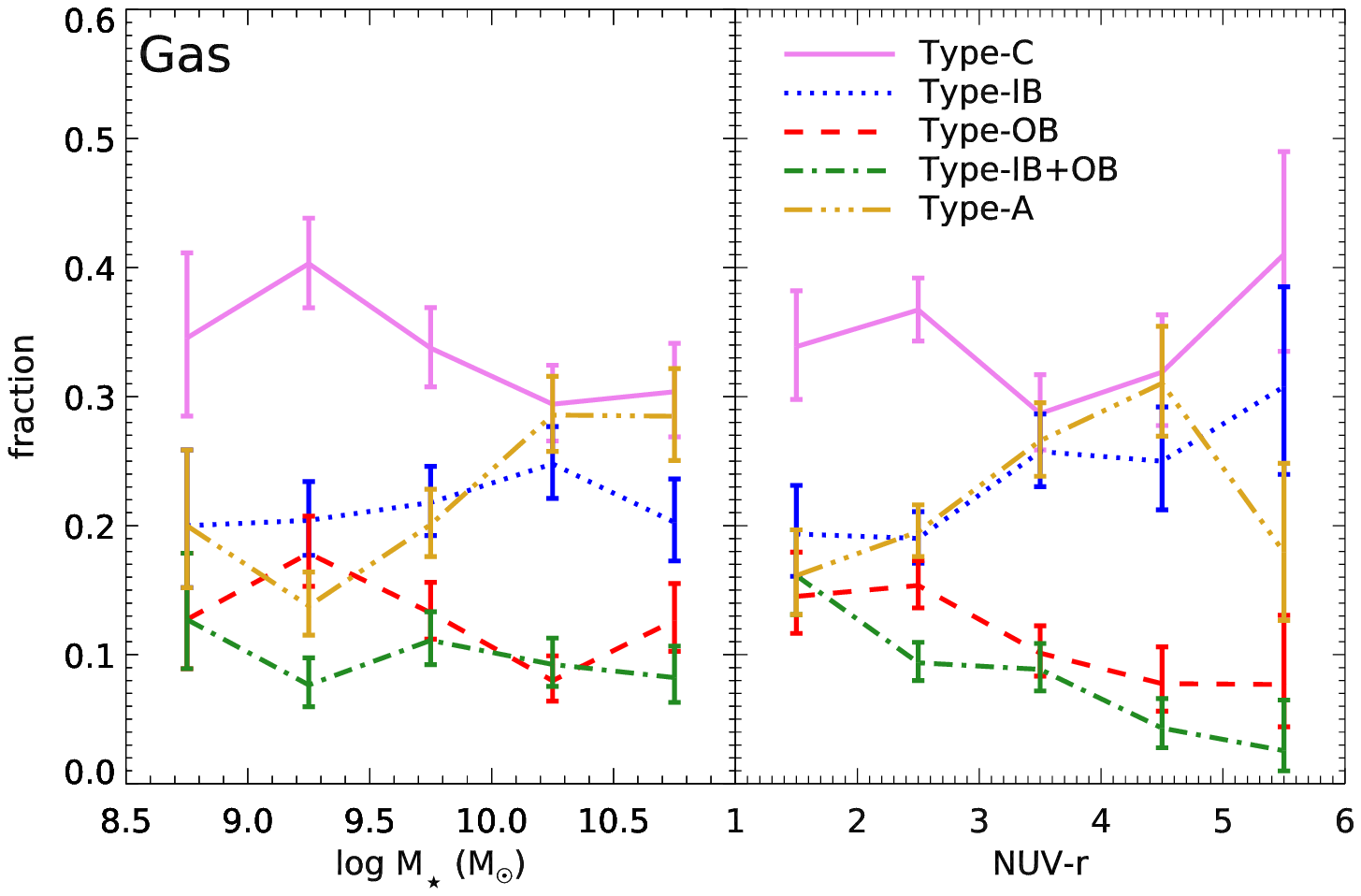}
\includegraphics[width=\columnwidth]{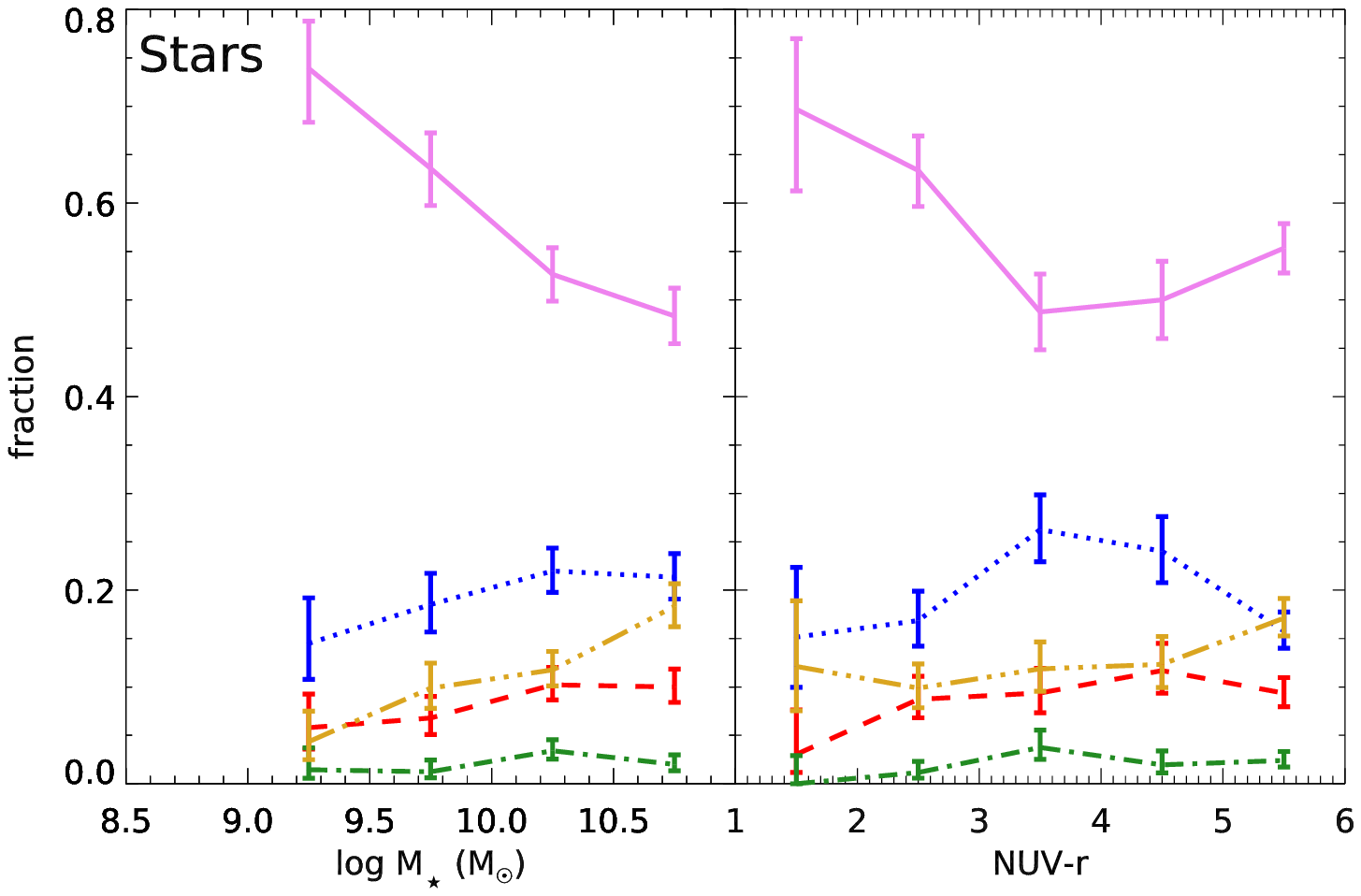}
\caption{Fraction of galaxies in different Radon profile classifications as a function of stellar mass and ${\rm NUV}-r$ color for gas (top) and stars (bottom). }
\label{fig:veltype_colormag_1d}
\end{figure}

\begin{figure*}
\includegraphics[width=2\columnwidth]{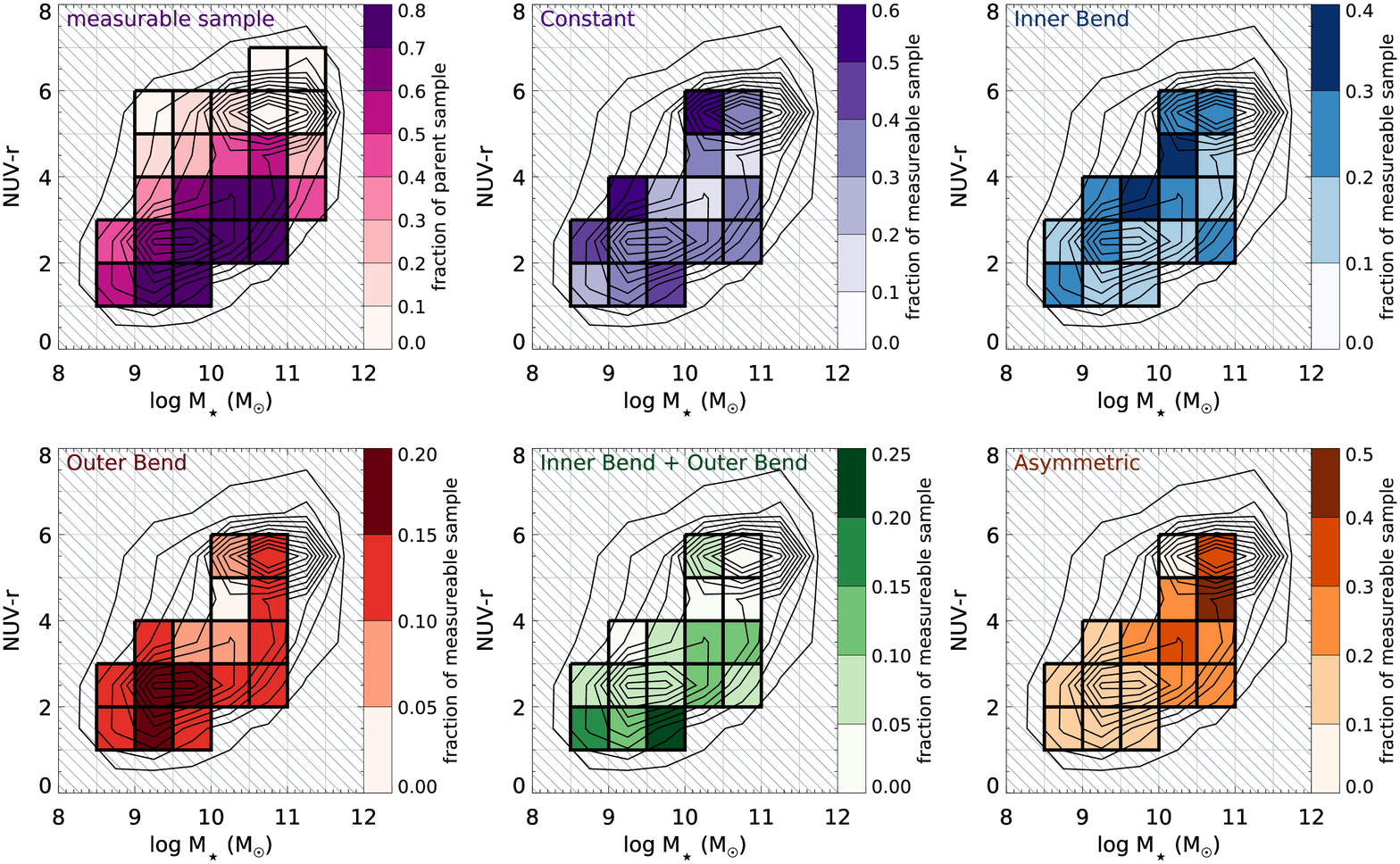}
\caption{Distribution of the frequency of each Radon profile classification for gas velocity fields in ${\rm NUV}-r$ versus $M_*$ parameter space. The upper left panel shows the fraction of galaxies which have radon profiles measurable out to at least $R_e$. All other panels show the fraction of galaxies in each category among those with measurable Radon profiles. The hashed regions are bins with fewer than $15$ galaxies in the denominator of the measured fraction, which are ignored due to the risk of large random errors. The dark gray contours show the distribution of the parent MaNGA sample.}
\label{fig:veltype_colormag_gas}
\includegraphics[width=2\columnwidth]{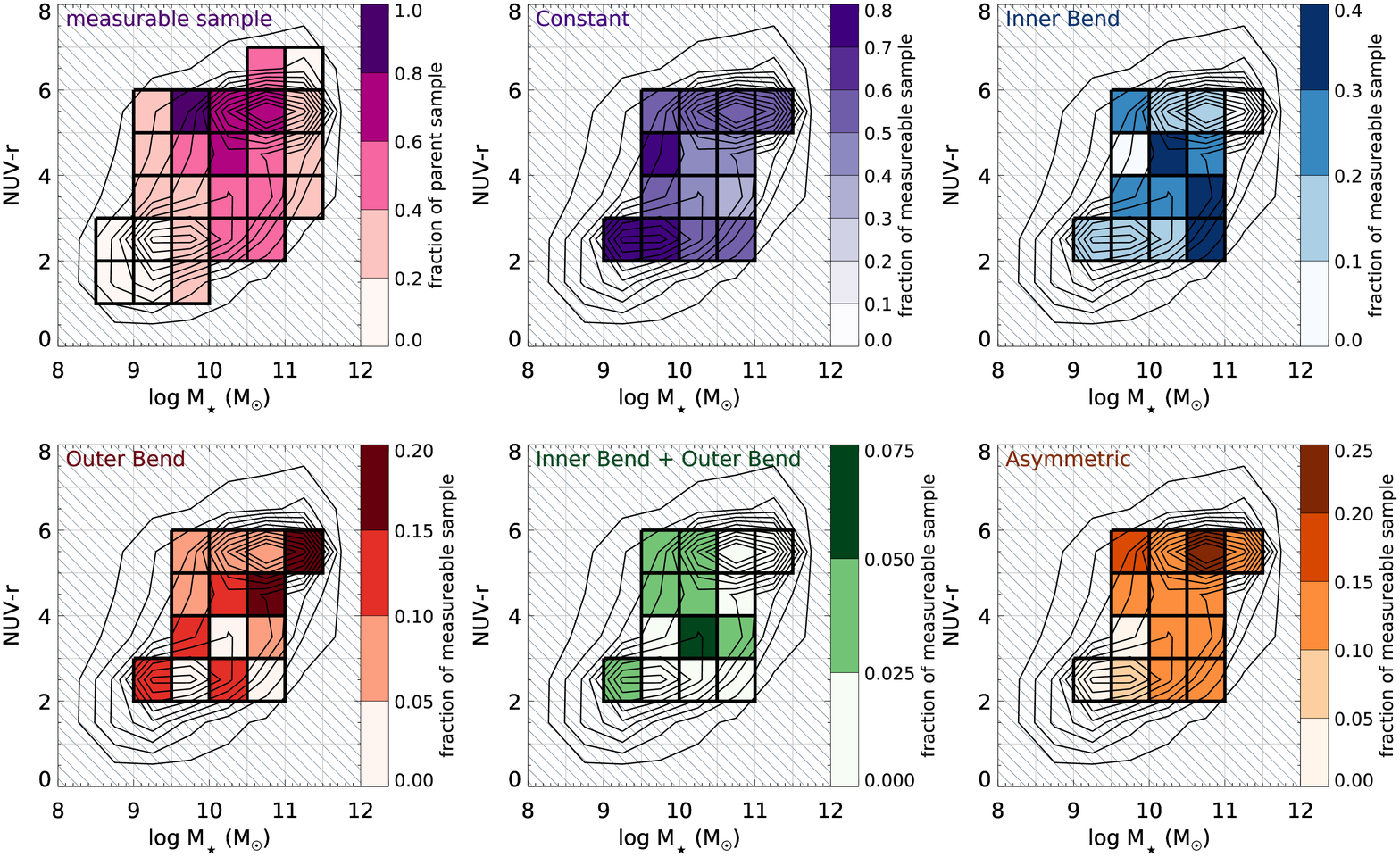}
\caption{Same as Fig.~\ref{fig:veltype_colormag_gas} but for stellar velocity fields.}
\label{fig:veltype_colormag_stars}
\end{figure*} 

To understand how different types of Radon profiles are distributed throughout the general galaxy population, we examine their frequency as a function of ${\rm NUV}-r$ color and stellar mass $M_*$ in Fig.~\ref{fig:veltype_colormag_1d}, which shows the frequency of each classification as a function of $M_*$ and ${\rm NUV}-r$ separately, and again in Fig.~\ref{fig:veltype_colormag_gas} and Fig.~\ref{fig:veltype_colormag_stars}, where we show the full 2D distribution of each classification. Stellar masses and photometric measurements come from the NSA, and bins in ${\rm NUV}-r$ are spaced by 1, while bins in $\log{M_*/M_{\odot}}$ are spaced by 0.5. The top-left panels in Fig.~\ref{fig:veltype_colormag_gas} and Fig.~\ref{fig:veltype_colormag_stars} show the fraction of galaxies out of the full parent sample where we were able to measure $\hat{\theta}(r)$ out to at least $R_e$, and the remaining panels show the fraction of galaxies which fall into each Radon profile classification normalized by the number of galaxies for which measurements could be made. The hashed regions indicate 2D bins where there were fewer than 15 galaxies with $\hat{\theta}(r)$ measurements, which we ignore to help minimize large random errors in the raster maps from small number statistics. For reference, the grey contours in each panel show the distribution of the full MaNGA sample for reference, regardless of whether we could measure $\hat{\theta}(r)$.

The distributions in Fig.~\ref{fig:veltype_colormag_1d} show that gas Radon profiles are largely independent of $M_*$ except for an increase in the frequency of Type-A profiles with increasing $M_*$.  More clear correlations are seen with respect color, with the Type-IB frequency showing a positive correlation with ${\rm NUV}-r$ and both Type-OB and Type-IB+OB frequencies showing negative correlations. Among stellar Radon profiles, the rates of Type-C and Type-A profiles are anti-correlated and correlated, respectively, with $M_*$, and no significant correlations are seen with respect to color.

Moving to the 2D distributions in Figs~\ref{fig:veltype_colormag_gas} and \ref{fig:veltype_colormag_stars}, we see that some types of Radon profiles prefer certain regions of color-mass space, although our selection effects make it difficult to disentangle the dependence on mass independent of color (and vice-versa), especially for gas velocity fields where our selection creates a correlation between these two quantities.  Among gas velocity fields, there is a slight excess of Type-OB profiles in the lower mass blue sequence, type-IB+OB profiles are almost exclusively found in the blue-sequence, and the fraction of Type-A profiles tends to increase with $M_*$ possibly becoming most frequent at redder colors. However, the low-mass red sequence is essentially untouched in these plots, making it difficult to determine whether mass or color (or both) is the primary driver of the observed trends.

The sampling of color-mass parameter space for stellar Radon profiles is more uniform.  We again see that the fraction of Type-C profiles decreases with $M_*$, although this trend appears to disappear in the red sequence above ${\rm NUV}-r\sim5$. We see some evidence that the fraction of Type-IB profiles increases with $M_*$ but only in the blue sequence below ${\rm NUV}-r\sim4$, and Type-A stellar Radon profiles appear particularly rare at low $M_*$ and ${\rm NUV}-r$.

Our sample size at this stage has limited our analysis to relatively coarse bin separations and with significant regions of parameter space ignored, so we reiterate that it is currently difficult to disentangle the relationship of Radon profile type with both stellar mass and color. The final MaNGA survey will be $\sim$4 times larger than the sample used in this study, at which point this analysis can be revisited to make a more robust determination of the link between kinematic behavior and these properties.

\section{Discussion}

\subsection{Physical Origin(s) of Different Radon Profiles}
\label{sec:origin}
In Section~\ref{sec:application} and Section~\ref{sec:census} we presented different types of regularly occurring patterns in Radon profiles, their frequencies, and their distribution within the galaxy population. The question remains as to what are the physical processes that cause the behavior presented in this work, and furthermore, whether the processes which drive kinematic behavior in the gaseous disk are completely distinct from the processes which drive the kinematic behavior of stellar disks. Although we do not definitively answer this question in this paper, we discuss which physical processes may be relevant, and what additional information can be used to test these ideas.

One of our key results is that roughly half of stellar velocity fields and two-thirds of gas velocity fields are inconsistent with uniform co-planar circular motion.  Similar results were seen in the CALIFA survey \citep{GarciaLorenzo15}. This high frequency of non-uniform {co-planar} circular motion implies that the processes which drive this behavior are either frequent or their effects long-lasting. Furthermore, the majority of galaxies with non-constant Radon profiles are still symmetric (Type-IB, Type-OB, and Type-IB+OB). Galaxies have been known to display such behavior for decades \citep[e.g.][]{Bosma81b} although thanks to large surveys like MaNGA we can obtain a stronger constraint on its frequency.  This behavior has typically been attributed to the presence of either (a) bar/oval distortions, or (b) warped disks. 

\subsubsection{Bar and Oval Distortions}

Given that bars are found in at least ${\sim}$30\% of galaxies \citep{Masters11}, they present a natural explanation for the large fraction of galaxies with {non-constant, but symmetric, Radon profiles}. We test this idea by crossmatching our sample with Galaxy Zoo \citep[GZ;][]{Willett13}, which provides bar classifications for roughly 85\% of our galaxies with Radon profile classifications. GZ relies on citizen scientists for bar identification, which are merged to estimate the probabilities that galaxies do or do not have bars ($p_{\rm bar}$ and $p_{\rm no bar}$), which are then adjusted to account for the redshift bias when identifying morphological features in observational data. Following previous work with GZ \citep{Skibba12,Masters12,Willett13,Kruk18}, we divide our sample into three classes: strong bars ($p_{\rm bar} > 0.5$), weak bars ($0.2 < p_{\rm bar} < 0.5$), and no bars ($p_{\rm bar} < 0.2$). The weak bar category likely contains some fraction of unbarred galaxies due to the increased difficulty identifying weaker bar structures.  \citet{Kruk18} find that $\sim$75\% of galaxies in this category show some signature of a bar.

\begin{figure*}
\includegraphics[width=\columnwidth]{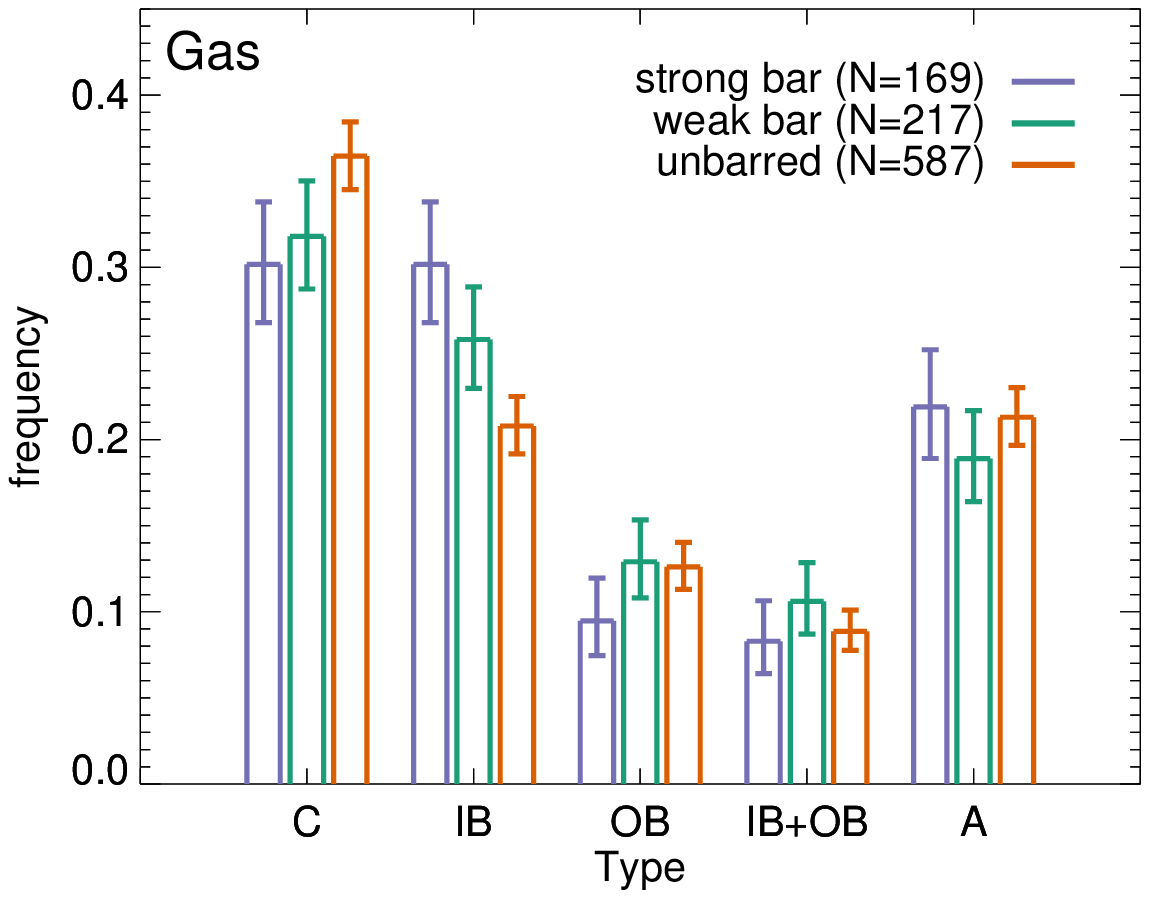}
\includegraphics[width=\columnwidth]{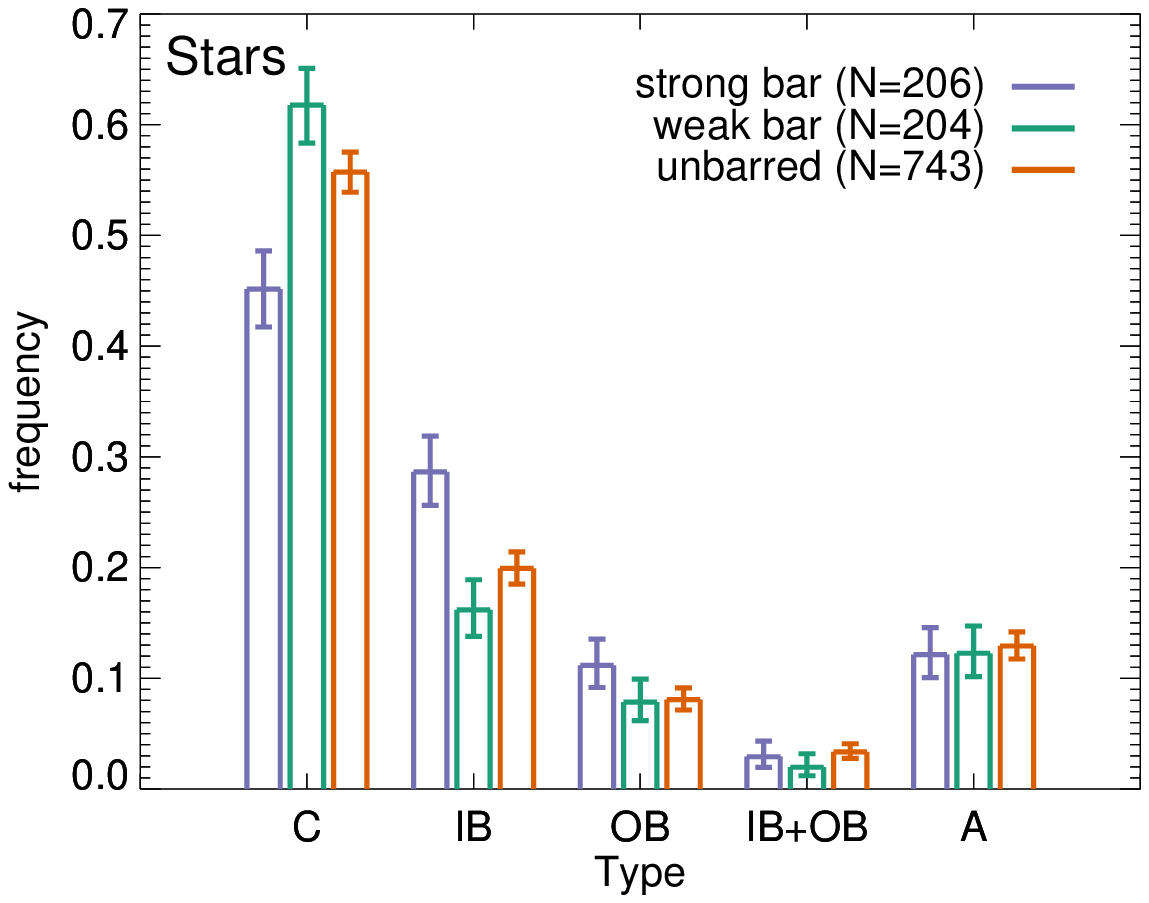}
\caption{Distribution of galaxies with strong, weak, and no bars into different Radon profile classifications for gas and stellar velocity fields.}
\label{fig:veltype_bar}
\end{figure*}

Fig.~\ref{fig:veltype_bar} shows the distribution of Radon profile classifications for gas and stellar velocity fields in strongly barred, weakly barred, and unbarred galaxies. Both gaseous and stellar Type-IB profiles are more frequent among strongly barred galaxies compared to unbarred galaxies, while unbarred galaxies have a clearly higher chance of being Type-C compared to strongly barred galaxies. For gas velocity fields, the weakly barred galaxies behave similar to the strongly barred galaxies, whereas for stellar velocity fields, the weakly barred galaxies behave more like the unbarred sample. Barred and unbarred galaxies show no significant difference in the rates in which they are classified as Type-OB, Type-IB+OB, and Type-A. These results suggest that bars are responsible for at least a subset of the Type-IB profiles we observe. The different behavior of gas and stellar Radon profiles also seems to suggest that weak bars can drive observable distortions in the gas velocity fields, but not the stellar velocity fields.  We have also checked whether Fig.~\ref{fig:veltype_bar} changes if we limit the sample to only those galaxies with classified Radon profiles for both gas and stars, but we find that it does not.

An additional line of evidence supporting Type-IB profiles as driven by bars comes from Fig.~\ref{fig:warp_fit_properties} which shows that the typical radius where the PA$_k$ shifts in Type-IB profiles typically occurs around $\sim0.5R_e$. \citet{DiazGarcia16} find that the ratio of bar length to $r$-band disk scale length is $r_{\rm bar}/h_r{\sim}1$ on average (with some variations of $\pm25\%$ depending on morphology). Assuming $R_e\sim1.7h_r$, we expect a typical bar length of ${\sim}0.6R_e$. In contrast, however, bar lengths measured in Galaxy Zoo \citep{Hoyle11} tend to be larger than the typical radius where PA$_k$ shifts. Recently, \citet{Kruk18} find the ratio of bar effective radius to galaxy effective radius is $\sim0.5$ on average, again in good agreement with the typical PA$_k$ shift radius of Type-IB profiles in Fig.~\ref{fig:warp_fit_properties}. A more detailed analysis of Radon profiles and bar structure is beyond the scope of this work, but these simple comparisons hint at a potential link between the location of transitions in PA$_k$ for Type-IB profiles and the length scales of bars.

There is still a very large fraction of barred galaxies classified as Type-C and a large fraction of unbarred galaxies classified as Type-IB, implying bars alone cannot drive Type-IB profiles, nor does the presence of a bar guarantee detectable distortions in the inner velocity field. There are a number of potential explanations for this result. First, it is possible for galaxies to experience gas inflows without the influence of a bar, perhaps instead due to tidal interactions \citep{Stark13}. Alternatively, if there is a bar it may be aligned with the rest of the disk such that there is no change in PA$_k$.  Additionally, the unbarred sample may be contaminated by weak bar features which GZ classifiers could not detect, or oval distortions which were not explicitly classified. Recently, \citet{Kruk18} found a large number of unbarred galaxies have clear oval distortions (which they refer to as inner lenses), and found that these galaxies had similar colors, masses, and sersic indices to barred galaxies, and clearly different colors from galaxies without bars or oval distortions. \citet{Kormendy79} argued that bars and oval distortions/lenses may be linked, with bars evolving into oval distortions over time. While the distinction between strong bars, weak bars, and oval distortions has historically been somewhat subjective, there is clear evidence that they represent the same underlying phenomenon --- elliptical structures with elongated orbits that may have a different PA$_k$ from the larger-scale disk --- and therefore should have similar effects on a velocity field. The fraction of galaxies with strong bars, weak bars, and oval distortions may be as high as $\sim$60\% \citep{Sellwood93,Knapen00,DiazGarcia16} so ignoring oval distortions may explain the large fraction of apparently unbarred galaxies with Type-IB Radon profiles.  

Although we lack visual classifications of oval distortions, a key indicator of an oval distortion is a non-orthogonal kinematic major and minor axis. We estimate major and minor kinematics PAs by measuring the value of $\theta$ where $R_{\rm AB}$ is minimized and maximized along all line segments running through the center of the galaxy ($\rho=0$). The actual calculation of these angles and their uncertainties follow the same procedures described in Section~\ref{sec:tracing_algorithm} and Section~\ref{sec:uncertainty}, but focusing only on the data at $\rho=0$.

\begin{figure*}
\includegraphics[width=\columnwidth]{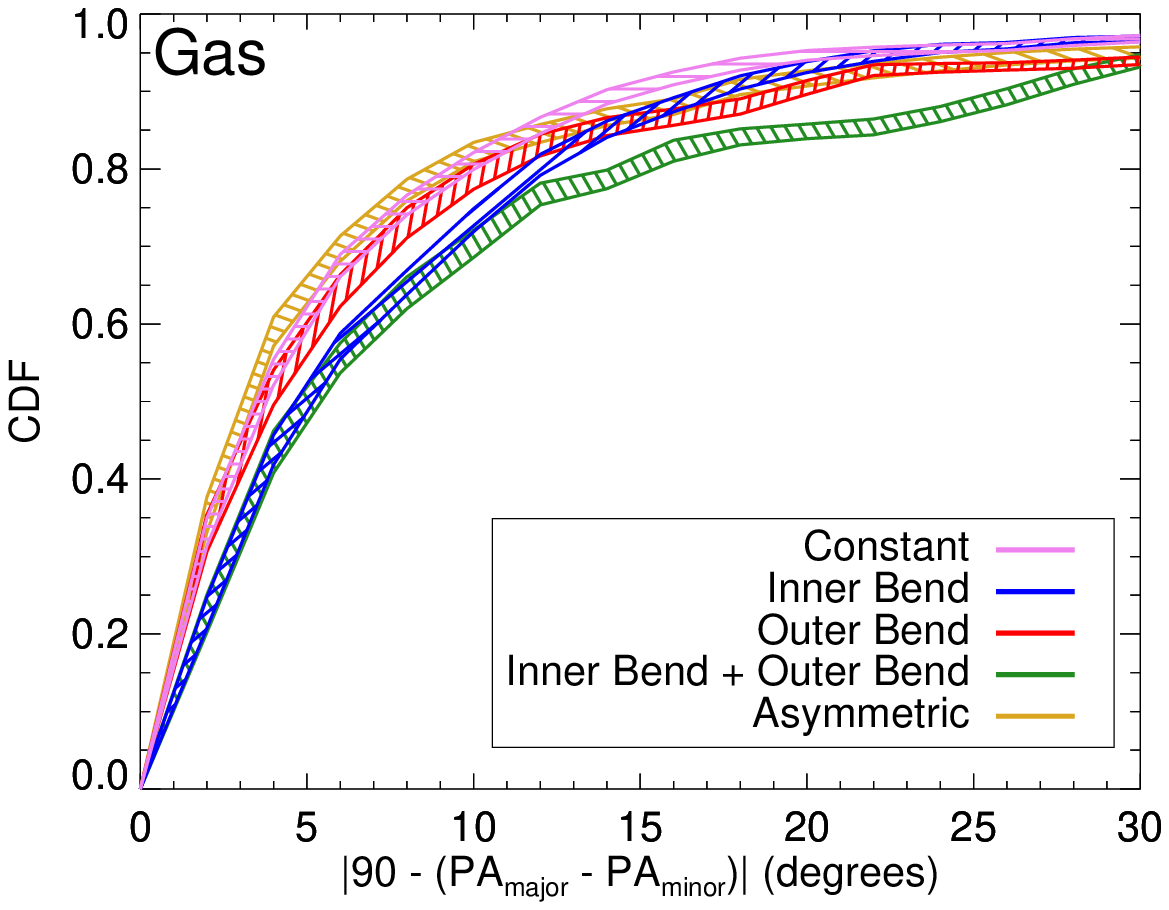}
\includegraphics[width=\columnwidth]{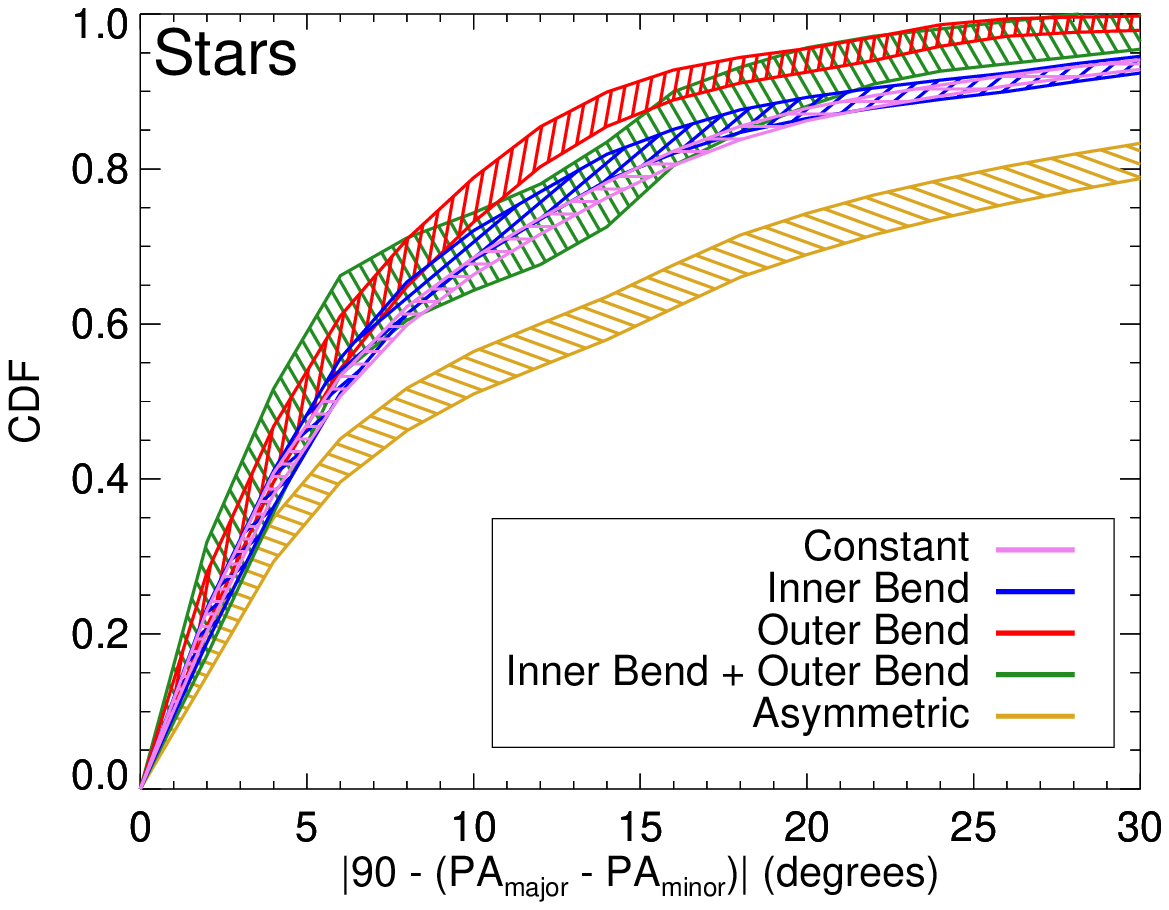}
\caption{Cumulative distribution functions (CDFs) of the opening angle between the major and minor PA$_k$ relative to 90$^{\circ}$. The width of each band corresponds to the $68\%$ confidence interval due to the uncertainty in the PA$_k$ measurements. The $x$-axis is truncated to $30^{\circ}$ to make the differences between the CDFs more clear.}
\label{fig:pa_major_minor}
\end{figure*}

Fig.~\ref{fig:pa_major_minor} shows the cumulative distribution functions (CDFs) of the difference between the major and minor axis PAs relative to 90$^{\circ}$ such that a value of $0^{\circ}$ corresponds to perfectly orthogonal major and minor axes, which we infer as true circular motion. For gas velocity fields, Type-IB and Type-IB+OB profiles show slightly more non-orthogonal motion at their centers compared to other types, implying they host more oval distortions.  This same behavior is not seen for stellar velocity fields, where the CDFs of both Type-C and Type-IB profiles imply similar rates of non-orthogonality.  The reason for this discrepancy is not clear, but one possible explanation may be the non-collisional nature of the stellar component. Both elongated and circular orbits will contribute to the observed velocity field at a given location, so any change in PA$_k$ induced by an oval distortion may be washed out by the underlying component with circular motion.  The gas, however, is collisional, so different kinematic components motion cannot coexist. Additional evidence in favor of this explanation can be seen in Fig.~\ref{fig:veltype_bar}, where the frequency of Type-IB gas Radon profiles in galaxies with weakly barred galaxies is similar to that of strongly barred galaxies, whereas the frequency of Type-IB stellar Radon profiles in galaxies with weakly barred galaxies is consistent with that of unbarred galaxies.  

\subsubsection{Kinematic Warps}

In addition to bars and oval distortions, kinematic warps may also drive symmetric radial variations in PA$_k$. Warps present a tempting explanation for Type-OB profiles for a number of reasons.  First, Type-OB profiles occur at similar frequencies independent of whether a galaxy has a strong bar (Fig.~\ref{fig:veltype_bar}), and the characteristic radius where $\thetahat$ changes in Type-OB profiles mostly occurs beyond the typical bar radius (Fig.~\ref{fig:warp_fit_properties}, \citealt{DiazGarcia16}), both of which suggest bars do not drive the observed distortions.  Second, Type-OB profiles typically show a smaller frequency of non-orthogonal major and minor kinematic PAs compared Type-IB and Type-IB+OB profiles, suggesting less influence from oval distortions. 

Disk warps themselves have a number of potential physical origins. Warps in gas disks may be the result of cosmic gas infall that is misaligned with the existing disk. If gaseous Type-OB profiles were indeed associated with such a scenario, we would expect them to cluster in the blue, low mass end in color-mass space which is typically populated by galaxies residing in low environmental densities where cool gas infall is expected to be most frequent \citep{Keres09}. While we do see an excess in this regime, it is very mild, and Type-OB profiles are seen well into the higher mass red-sequence.  However, warped gas disks are also an expected phenomenon among red (typically early-type) galaxies; $\sim30\%$ of fast-rotator early-type galaxies have ionized gas that is misaligned, suggesting an external origin \citep{Davis11}, and a warp will naturally result from a misaligned gas disk as the stellar disk torques the gas into alignment from the inside out \citep{VanDeVoort15}. In such a case, the gas disk may form from cosmological accretion or from the debris left over from recent interactions \citep{Davis11,Serra12,Lagos15}. The relatively high rate of gas Type-OB profiles in the high-mass red regime may itself be driven by a selection effect under this scenario; gas disks in these galaxies tend to have an external origin, and a warp is a natural result of acquiring a new gas disk.

Warps in stellar disks may result from multiple stellar components, suggesting they result from mergers. Several studies have argued that early-type galaxies are assembled through dry mergers, and the majority have had a major merger in their recent history \citep{VanDokkum05,Bell06} which naturally explains distinct kinematic components in the remnant. Since early-type galaxies dominate the galaxy population in the high-mass red sequence \citep[e.g.][]{Nakamura03}, we expect the majority of stellar Type-OB profiles to fall in this regime if they do in fact reflect multiple component stellar systems, which is hinted at in Fig.~\ref{fig:veltype_colormag_stars}. It has also been argued that more distant tidal interactions can sustain galaxy warps \citep[e.g.][]{Weinberg06}. Combining environmental information with estimates of gas content and metallicity can help constrain the importance of cosmic accretion versus merging and tidal interactions in driving gas and stellar warps.

\subsubsection{Tidal Interactions}

Tidal interactions present a natural explanation for asymmetric Radon profiles. \citet{GarciaLorenzo15} find a similar rate of asymmetry in ionized gas velocity fields in CALIFA and find most cases can be explained by tidal interactions. The increasing rate of gas asymmetries at higher stellar masses is qualitatively consistent with the expected trend from merger rate \citep{Xu04,Patton08,Casteels14,RodriguezGomez15}, although properly converting the rate of asymmetries into a merger rate requires both completeness corrections and estimates of the timescale over which asymmetry is visible, both of which we have not attempted here. Interestingly, although they obtain a similar frequency of asymmetric gas velocity fields (using Kinemetry), \citet{Bloom17} find the rate of kinematic asymmetry actually decreases with increasing stellar mass. However, their result is largely driven by galaxies with $M_*<10^{8.5}\,M_{\odot}$ that we do not probe in our study. It is also notable that gas velocity fields are ${\sim}2$ times more likely to be asymmetric than stellar velocity fields, and stellar Type-A profiles show no clear mass dependence. However, such behavior is not necessarily surprising given that gas is more sensitive to tidal perturbations and tends to reflect past merger events for significantly longer timescales than stars \citep{Holwerda11}.

Tidal interactions may not be the only cause of asymmetry, however. \citet{Bournaud05} argued that the rate of lopsided disks cannot be explained by interactions alone, and that cosmic gas infall provides an additional driver (a caveat of this comparison is that \citet{Bournaud05} was largely  focusing on morphological, not kinematic, lopsidedness).  Cosmic gas infall explanation seems an unlikely explanation at our high mass end (where asymmetries are most common) given that rapid cool gas infall should be most prevalent among lower stellar mass blue sequence galaxies \citep{Keres09}. Nonetheless, analyzing the environmental properties, gas fractions and metallicities of asymmetric galaxies will help test the merger vs. gas infall scenarios. Asymmetry may also arise due to more turbulent support, possibly from disk instabilities and/or feedback from massive star forming clumps, which again for our sample is most likely to impact galaxies at lower stellar mass.  Follow-up work using spatially resolved maps of star formation and velocity dispersion may help constrain this possibility.

The variety of kinematic behavior within the galaxy population can arise from a number of different processes.  Although we have identified different types of characteristic behavior, assessed their frequency, and speculated to some degree as to their origins, a more detailed analysis combining information about key properties like environment, star formation histories, gas content, and metallicity will be needed to obtain a better understanding of what physical processes galaxy kinematics are reflecting.  This more in-depth analysis will be the focus of future work.

\subsection{Advantages and Disadvantages of the Radon Transform Over Other Methods}

There are multiple tools currently available to analyze velocity fields, many of which can also assess any radial variability in PA$_k$. This fact naturally raises the question of why would one want to use the Radon transform over these other approaches.  

{We argue that if one is interested in determining the orientation and/or regularity of the velocity field, the Radon transform provides a straightforward means of obtaining this information. The Radon transform simplifies the analysis by collapsing the velocity field into a form where changes in PA$_k$ are the main focus, essentially ignoring inclination and kinematic center and making the analysis of PA$_k$ more robust. Such an approach particularly advantageous for low S/N data, where simultaneously determining properties like disk center and inclination in addition to position angle, can yield highly unstable results \mbox{\citep[see e.g.,][]{Bloom17}}. That being said, other approaches such as Kinemetry \mbox{\citep{Krajnovic06}} or tilted-ring fitting \citep[e.g.,][]{Begeman87} typically allow one to fix these other properties so that radial profiles of PA$_k$ can be extracted more robustly. What distinguishes the Radon transform from other methods is that it is non-parametric; at its heart the velocity field is not tied to an underlying model, and we simply transform the data to enable easy assessment of PA$_k$\footnote{{Although we do extract $\thetahat(\rho)$ ($\rm PA_k$) using a Von Mises function (Section~\ref{sec:tracing_algorithm}) and classify galaxies into different categories using several analytical functions (Section~\ref{sec:classification}), neither of these steps involve assuming any functional form for the underlying velocity field.}}. This simplicity makes the Radon transform easily automated and ideal for characterizing large samples.}


{Although the simplicity of the Radon transform can be a major advantage, for certain purposes its simplicity may be a major disadvantage. If one is interested in determining of additional properties of the velocity field (e.g., disk inclination) or one wants to model different kinematic components (e.g., radial flows; \citealt{Spekkens07}), alternative tools must be used. However, the output from the Radon transform may still be a useful input to these other tools, allowing them to run with PA$_k$ already well-constrained, thus simplifying any additional fitting and parameter estimation.  Nonetheless, if one is simply interested in the orientation and regularity of a galaxy's velocity field, the Radon transform provides a straightforward, non-parametric means of attaining this information.}

\section{Conclusions}
\label{sec:conclusion}

We have illustrated how the Radon transform can be used to quantify the mean kinematic PAs of velocity fields, as well as any radial variations or asymmetries.  This technique provides a simple, non-parametric means of assessing the orientation and regularity of velocity fields, and can be particularly useful for ongoing and upcoming IFU and radio interferometric surveys which yield samples of thousands of galaxies.

We have applied this technique to gas and stellar velocity fields from the MaNGA IFU survey and measured ``Radon profiles" (which correspond to radial profiles of kinematic position angle) for the first $\sim$2800 MaNGA galaxies.  We place galaxies into five different classes based on their Radon profiles: Type-C (constant) profiles where $\thetahat$ shows no significant radial variation, Type-IB (inner bend) profiles where $\thetahat$ changes immediately starting at the center typically out to $\sim R_e/2$, Type-OB (outer bend) profiles where the PA$_k$ shifts to a new value at some radius away from the center, Type-IB+OB (inner bend + outer bend) profiles which show a combination of Type-IB and Type-OB profiles, and Type-A (asymmetric) profiles where $\thetahat$ show significantly different behavior on either side of the Radon profile corresponding to the approaching and receding side of the galaxy. Although these different classifications are phenomenological, some are better suited to capture specific processes, e.g., Type-OB may best capture outer disk warps, while Type-IB may best capture bars or oval distortions in the inner disk.

Our key findings are:
\begin{itemize}
\item Approximately half of stellar velocity fields and two-thirds of gas velocity fields show non-constant Radon profiles.  In the majority of these cases, the variation is symmetric about the center (Section~\ref{sec:census}, Table~\ref{table:warp_class_fractions}).  
\item The distribution of characteristic radii where the kinematic PAs change are very similar for gas and stellar velocity fields (Section~\ref{sec:structure}, Fig.~\ref{fig:warp_fit_properties}).  However, the actual Radon profile classifications of gas and stars are largely independent of one another (Section~\ref{sec:star_gas_agreement}, Fig.~\ref{fig:stars_vs_gas_type}).   
\item Some Radon profile classifications cluster in certain regions within NUV$-r$ versus stellar mass space, and the distribution of stellar and gaseous Radon profile classifications are largely independent of one another. Among gas velocity fields, Type-OB profiles are more common in the low-mass blue sequence, Type-IB+OB profiles are more common in the blue sequence, and Type-A profiles are more common at higher stellar mass.  Among stellar velocity fields, Type-C profiles appear less frequently at higher mass, while Type-IB profiles become more frequent, although these trends may not hold on the red sequence. (Section~\ref{sec:color_mass}, Fig.~\ref{fig:veltype_colormag_gas}, Fig.~\ref{fig:veltype_colormag_stars}).
\item Barred galaxies are more likely to be associated with Type-IB profiles compared to unbarred galaxies, but there are still large fractions of barred galaxies in other categories, and unbarred galaxies classified as Type-IB.  Gaseous Type-IB and Type-IB+OB profiles show larger fractions of non-orthogonal major and minor axis kinematic PAs, suggesting their behavior may be associated with oval distortions (Section~\ref{sec:origin}, Fig.~\ref{fig:veltype_bar}, Fig.~\ref{fig:pa_major_minor}).  
\item Type-OB profiles show no clear dependence on the presence of a bar and show more orthogonal major and minor axis kinematic PAs compared to Type-IB and Type-IB+OB profiles.  These trends suggest Type-OB profiles are more likely to be associated with warped disks as opposed to bar or oval distortions (Section~\ref{sec:origin}, Fig.~\ref{fig:veltype_bar}, Fig.~\ref{fig:pa_major_minor}).
\end{itemize}

Future work will explore how different Radon profiles relate to other galaxy properties, specifically their environments, star formation histories, metal abundances, and gas content.  This more in-depth analysis will greatly enhance our ability to constrain the physical mechanisms (e.g., gas infall, tidal interactions, secular processes) which drive the observed kinematic behavior throughout the galaxy population. The MaNGA survey itself provides much of the information needed to understand the internal galaxy properties, and the broader SDSS legacy redshift survey enables a robust understanding of environment for much of the sample.  Lastly, the ongoing MaNGA-HI follow-up survey (Masters et al. in prep) will provide a large inventory of atomic hydrogen gas masses.

\section*{Acknowledgements}
We are very grateful to our referee for their review which helped improve this paper. We also thank Preethi Nair for useful discussions regarding the role of bars. AW acknowledges support from a Leverhulme Trust Early Career Fellowship. MAB acknowledges support from NSF AST-1517006. DB acknowledges support from grant RSCF-14-50-00043. This publication uses data generated via the Zooniverse.org platform, development of is which funded by generous support, including a Global Impact Award from Google, and by a grant from the Alfred P. Sloan Foundation.  

Funding for the Sloan Digital Sky Survey IV has been provided by the Alfred P. Sloan Foundation, the U.S. Department of Energy Office of Science, and the Participating Institutions. SDSS-IV acknowledges
support and resources from the Center for High-Performance Computing at
the University of Utah. The SDSS web site is www.sdss.org.

SDSS-IV is managed by the Astrophysical Research Consortium for the 
Participating Institutions of the SDSS Collaboration including the 
Brazilian Participation Group, the Carnegie Institution for Science, 
Carnegie Mellon University, the Chilean Participation Group, the French Participation Group, Harvard-Smithsonian Center for Astrophysics, 
Instituto de Astrof\'isica de Canarias, The Johns Hopkins University, 
Kavli Institute for the Physics and Mathematics of the Universe (IPMU) / 
University of Tokyo, the Korean Participation Group, Lawrence Berkeley National Laboratory, 
Leibniz Institut f\"ur Astrophysik Potsdam (AIP),  
Max-Planck-Institut f\"ur Astronomie (MPIA Heidelberg), 
Max-Planck-Institut f\"ur Astrophysik (MPA Garching), 
Max-Planck-Institut f\"ur Extraterrestrische Physik (MPE), 
National Astronomical Observatories of China, New Mexico State University, 
New York University, University of Notre Dame, 
Observat\'ario Nacional / MCTI, The Ohio State University, 
Pennsylvania State University, Shanghai Astronomical Observatory, 
United Kingdom Participation Group,
Universidad Nacional Aut\'onoma de M\'exico, University of Arizona, 
University of Colorado Boulder, University of Oxford, University of Portsmouth, 
University of Utah, University of Virginia, University of Washington, University of Wisconsin, 
Vanderbilt University, and Yale University.




\bibliographystyle{mnras}
\bibliography{mybib} 








\bsp	
\label{lastpage}
\end{document}